\documentclass[twocolumn,twocolappendix]{aastex63}
\usepackage{amsmath}
\usepackage{multirow}
\usepackage{subfigure}
\usepackage{inputenc}
\newcommand{\msun}{\ensuremath{\,\rm{M}_\odot}}
\newcommand{\msunyr}{\ensuremath{\rm{M}_\odot\,\rm{yr}^{-1}}}

\newcommand{\ergs}{\ensuremath{\rm{erg/s}}}

\newcommand{\mint}{\ensuremath{\rm{M_{init}}}}

\newcommand{\crate}{\ensuremath{^{12}\rm{C}\left(\alpha,\gamma\right)^{16}\!\rm{O}}}

\newcommand{\code}[1]{\texttt{#1}}
\newcommand{\mesa}{\code{MESA}}
\newcommand{\MESA}{\code{MESA}}

\newcommand{\REACLIB}{\code{REACLIB}}

\newcommand{\kms}{{\mathrm{km\ s^{-1}}}}

\newcommand{\nuclei}[2]{\ensuremath{\mathrm{^{#1}#2}}}

\newcommand{\hydrogen}[1][1]{\nuclei{#1}{H}}
\newcommand{\helium}[1][4]{\nuclei{#1}{He}}
\newcommand{\lithium}[1][7]{\nuclei{#1}{Li}}
\newcommand{\beryllium}[1][9]{\nuclei{#1}{Be}}

\newcommand{\carbon}[1][12]{\nuclei{#1}{C}}
\newcommand{\nitrogen}[1][14]{\nuclei{#1}{N}}
\newcommand{\oxygen}[1][16]{\nuclei{#1}{O}}
\newcommand{\fluorine}[1][19]{\nuclei{#1}{F}}
\newcommand{\neon}[1][20]{\nuclei{#1}{Ne}}
\newcommand{\sodium}[1][23]{\nuclei{#1}{Na}}
\newcommand{\magnesium}[1][24]{\nuclei{#1}{Mg}}
\newcommand{\aluminum}[1][27]{\nuclei{#1}{Al}}

\newcommand{\chlorine}[1][35]{\nuclei{#1}{Cl}}

\newcommand{\potassium}[1][39]{\nuclei{#1}{K}}
\newcommand{\calcium}[1][40]{\nuclei{#1}{Ca}}
\newcommand{\scandium}[1][45]{\nuclei{#1}{Sc}}

\newcommand{\vanadium}[1][51]{\nuclei{#1}{V}}
\newcommand{\chromium}[1][52]{\nuclei{#1}{Cr}}
\newcommand{\manganese}[1][55]{\nuclei{#1}{Mn}}
\newcommand{\iron}[1][56]{\nuclei{#1}{Fe}}
\newcommand{\cobalt}[1][59]{\nuclei{#1}{Co}}
\newcommand{\nickel}[1][58]{\nuclei{#1}{Ni}}
\newcommand{\copper}[1][63]{\nuclei{#1}{Cu}}
\newcommand{\zinc}[1][64]{\nuclei{#1}{Zn}}

\begin{document} 

\title{Nucleosynthesis of binary-stripped stars.}

\author[0000-0003-3441-7624]{R.~Farmer}
\email{rfarmer@mpa-garching.mpg.de}
\affiliation{Max-Planck-Institut f{\"u}r Astrophysik, Karl-Schwarzschild-Straße 1, 85741 Garching, Germany}  

\author[0000-0003-1009-5691]{E.~Laplace}
\affiliation{Heidelberger Institut f{\"u}r Theoretische Studien, Schloss-Wolfsbrunnenweg 35, 69118
Heidelberg, Germany}
\affiliation{Anton Pannekoek Institute for Astronomy and GRAPPA, University of Amsterdam, 
NL-1090 GE Amsterdam, The Netherlands}

\author[0000-0002-9911-8767]{Jing-ze Ma}
\affiliation{Max-Planck-Institut f{\"u}r Astrophysik, Karl-Schwarzschild-Straße 1, 85741 Garching, Germany}

\author[0000-0001-9336-2825]{S.~E.~de~Mink}
\affiliation{Max-Planck-Institut f{\"u}r Astrophysik, Karl-Schwarzschild-Straße 1, 85741 Garching, Germany}
\affiliation{Anton Pannekoek Institute for Astronomy and GRAPPA, University of Amsterdam, 
NL-1090 GE Amsterdam, The Netherlands}

\author[0000-0001-7969-1569]{S.~Justham}
\affiliation{School of Astronomy \& Space Science, University of the Chinese Academy of Sciences, 
Beijing 100012, China}
\affiliation{National Astronomical Observatories, Chinese Academy of Sciences, Beijing 100012, China}
\affiliation{Anton Pannekoek Institute for Astronomy and GRAPPA, University of Amsterdam, NL-1090 
GE Amsterdam, The Netherlands}
\affiliation{Max-Planck-Institut f{\"u}r Astrophysik, Karl-Schwarzschild-Straße 1, 85741 Garching, Germany} 

\date{\today}

\begin{abstract}

The cosmic origin of the elements, the fundamental chemical building blocks of the Universe, is still uncertain. 
Binary interactions play a key role in the evolution of many massive stars, yet their impact on chemical yields is poorly understood.
Using the MESA stellar evolution code we predict the chemical yields ejected in wind mass loss and the supernovae of single and binary-stripped stars. We do this with a large 162 isotope nuclear network at solar-metallicity.
We find that binary-stripped stars are more effective producers of the elements than single stars, due to their increased mass loss and an increased chance to eject their envelopes during a supernova. 
This increased production by binaries varies across the periodic table, with \fluorine[] and \potassium[] being more significantly produced by binary-stripped stars than single stars.
We find that the \carbon[12]/\carbon[13] could be used as an indicator of the conservativeness of mass transfer, as \carbon[13] is preferentially ejected during mass transfer while \carbon[12] is preferentially ejected during wind mass loss.
We identify a number of gamma-ray emitting radioactive isotopes that may be used to help constrain progenitor and explosion models of core-collapse supernovae with next-generation gamma-ray detectors. For single stars we find \vanadium[44] and \manganese[52] are strong probes of the explosion model, while for binary-stripped stars it is \chromium[48].
Our findings highlight that binary-stripped stars are not equivalent to two single stars and that detailed stellar modelling is needed to predict their final nucleosynthetic yields.

\end{abstract}

%
\section{Introduction}\label{sec:intro}  

The origin of the elements is an unsolved problem \citep{arnould99,jose11,diehl22}. Massive stars play a role in the formation of the 
elements through their ejection of processed material in their winds \citep{kudritzki00,vink:01,crowther07,smith14} and the ejecta from
core-collapse (CC) supernovae \citep{maeder92,rauscher02,heger:03}. 

Most massive stars have been found to be born with at least one companion \citep{Abt1983,mason:09}, and the majority are expected to undergo at least one phase of mass transfer \citep{sana:12,moe:17}. During mass transfer the majority of the hydrogen-rich envelope of the star \citep{kippenhahn:67,yoon:10,gotberg17}
may be removed and some fraction of it accreted onto its companion. This mass transfer alters both stars' subsequent evolution \citep{podsiadlowski92,brown96,wellstein99,eldridge08,langer12,demink:13} and final chemical element production \citep{braun95,laplace21,farmer21}.
Binarity has also been used to explain the composition of carbon and s-element enhanced metal-poor (CEMP-s) stars, by requiring efficient wind mass transfer in the system \citep{abate2015a}.

Explosive nucleosynthesis, during core-collapse (CC) supernova (SN), in massive stars has been well studied for single star objects \citep{woosley93b,heger01,heger02,heger02b,chieffi04,woosley07,pignatari16,limongi18} but has been less well studied for binary progenitors (\citealt{braun95b,brinkman19,brinkman21}; see, however, \citealt{izzard04,izzard06})
The primary uncertainty in the nucleosynethic yields of CC supernovae is whether the star explodes and forms a neutron star (NS), ejecting its envelope, or collapses into a black hole (BH), where the envelope accretes onto the BH. The boundary between these fates is uncertain \citep{heger:03,ertl:15,boccioli22}, with binary interactions adding additional complexity \citep{timmes96,brown01,podsiadlowski04}.

Chemical enrichment of galaxies provides a direct test of both the formation of stars and their evolution \citep{tinsley80,pagel97}, though these have also still mostly relied upon single-star models (\citealt{nomoto13,kobayashi20}; see however \citealt{dedonder04}). Wide-field Spectroscopic surveys (such as APOGEE \citealt{majewski17}), combined with Gaia \citep{gaia21,gaia2022} are revealing the chemical composition and evolution of the Milky Way \citep{perryman01,lian22}. 

Massive stars are not the only source of chemical enrichment in the Universe \citep{burbidge57}.
Other sources include the Big Bang \citep{peebles66,wagoner67}, asymptotic giant branch (AGB) stars \citep{vandenhoek97,herwig05,karakas10}, type Ia SNe \citep{woosley94,nomoto97,iwamoto99,eitner22}, and double neutron star mergers \citep{freiburghaus99,arnould07,kasen17,pian17,tanvir17}. Each source leaves a different chemical fingerprint on the composition of the Universe. Each source also contributes different elements and at different times \citep{carigi05,cescutti09}. For instance, massive stars evolve rapidly and can enrich the Universe in a few million years after star formation occurs \citep{franchini20}. While AGB \citep{nissen14}, SN Ia \citep{leung20}, and double neutron star mergers \citep{dominik:12,dominik:13} take much longer to enrich the Universe, due to the slower evolution of their progenitor stars and/or long gravitational-wave inspiral times before merger.

In Section \ref{sec:meth} we discuss our evolution of single and binary stars, and the properties of our explosion models. In Section \ref{sec:yields} we discuss the total chemical yields from different mass-loss processes, while Section \ref{sec:res_winds} breaks down the pre-supernovae yields and section \ref{sec:res_sn} discuss the post-supernovae yields. Section \ref{sec:caveats} discusses several caveats with our results. Finally we discuss and conclude our results in Sections \ref{sec:discus} and \ref{sec:conc}.

\section{Method}\label{sec:meth}

This work builds upon the work in \citet{farmer21}, which explored the chemical yields of 
\carbon{}, to extend the discussion to a larger set of isotopes (162) up to \zinc[64]. We evolve a grid of 
single stars and binary-stripped stars
using the MESA stellar evolution code  \citep[version 12115,][]{paxton:11,paxton:13,paxton:15,paxton:18,paxton:19,jermyn2022}. These stars are evolved from the
zero-age main sequence to core collapse and then through their supernovae until shock breakout.
Our single stars and the primary (initially most massive star) in the binary have initial 
masses between $\mint=11$ -- $45\msun{}$. For
binary stars we set the initial period to be
between 38\textendash300 days and we use a mass ratio of $\rm{M_2/M_1}=0.8$ to set each secondary
star's mass.
The orbital period chosen for each system ensures that all the binary stars undergo case B mass transfer \citep{paczynski67b,vandenheuvel69}.
We follow the mass transfer from the primary star in the binary onto the companion assuming fully conservative mass transfer during Roche-lobe overflow (RLOF). 

All models
are computed with an initial solar metallicity of Z=0.0142 and are non-rotating. 
The single stars and binary stars 
are evolved with an initial Y=0.2684 \citep[$\rm{Y=2Z+0.24,}$][]{pols95,tout96}.
All stars are set to have the same initial composition profile, and is based on the solar composition of \citet{grevesse:98}. Additional physics options are specified in Appendix \ref{sec:other_phys}. Four models did not reach core collapse and are excluded from this analysis, these are the binary-stripped star models with initial masses of 24, 25, 28, and 29\msun{}.

Inlists with all input parameters, data tables, and auxiliary data are made available online at \url{https://doi.org/10.5281/zenodo.5929870}. A sample of the data available is shown in Appendix \ref{sec:data_table}. While we have selected theoretically and observationally motivated assumptions for our input physics, we have not performed a systematic calibration of our models as we are interested in the \textit{relative} differences between the chemical yields of  binary-stripped and single stars. 

To simplify our models we evolve the primary star in a binary system with a point-mass companion
until the end of core helium burning. At this point we remove the companion and continue the evolution of the binary until core collapse, assuming these systems will not interact again \citep{laplace20}. Assuming that
the Roche-lobe radius does not decrease after core helium burning and comparing that with the maximum radial extent of our binary stars post core helium burning, we find that this is a good approximation except for the $\mint=22$ and $27\msun$ models which would exceed their Roche-lobe radii (measured at core helium depletion), and the $\mint=23\msun$ which reaches 80\% of its Roche-lobe radius (measured at core helium depletion). For all other binary-stripped stars the maximum radius is $<10\%$ of their Roche-lobe radii (measured at core helium depletion). Whether these stars would interact again, depends on detailed modelling of the accretor, as any mass it loses via winds would widen the orbit preventing further mass transfer phases.

Our choice of physics and and modelling assumptions were described in \citet{farmer21} and builds upon the work
presented in \citet{laplace20,laplace21}. We update our choice of convective overshoot in ``metal'' burning regions (regions burning elements heavier than helium) to an exponential profile (instead of a step overshoot profile). For that overshooting we assume a value of $f=0.03$ and $f_0=0.001$ above the metal burning regions and $f=0.003$ and $f_0=0.0001$ below the metal burning regions \citep{jones17},
where $f$ is the extent of the overshoot region in pressure scale heights and $f_0$ is the point inside the convection zone where the strength of convective overshoot begins exponentially decaying (in pressure scale heights). For the other overshoot regions we assume a step overshoot profile with $f=0.385$ and $f_0=0.05$ \citep{brott11}. We use the wind prescriptions of \citet{dejager88,nugis:00,vink:01} (``Dutch''), with a wind efficiency of 1.0. We include semiconvective mixing with a $\alpha_{\rm{semi}}=1.0$ and do not include thermohaline mixing \citep{farmer16}. We also include MLT++ to improve the numerical stability of the hydrogen envelopes \citep{paxton:13}.

The larger nuclear network we adopt, compared to \citet{farmer21}, is not only necessary for our yield predictions but is also important to capture important physics associated with the evolution up to core collapse \citep{farmer16}.
We use a 162 isotope nuclear network which contains
\MESA's \texttt{mesa\_161.net}\footnote{This network misses several stable isotopes, namely \copper[65], \zinc[66-68], and \zinc[70]. } network plus \calcium[47], for all evolutionary stages including the supernova phase. This network was chosen as a balance between larger networks (to capture more physics) and being computationally feasible. The addition of \calcium[47] was so that we could cover all isotopes that were found to be of interest for next generation gamma-ray observatories
\citep{timmes19,andrews20}.

We define the helium core mass of the star as where the helium mass fraction, $
\rm{X_{He}}>0.1$ and the hydrogen mass fraction (at the same mass coordinate) is $\rm{X_{H}}<0.01$. Finally, we define core collapse to occur when the inner regions of the star infall at $300\kms$.

We define the yield of an isotope as \citep{karakas16}:

\begin{equation}\label{eq:yields}
     \rm{Yield} = \sum_T \Delta M_T \times \left(X_j - X_{j,init}\right)
\end{equation}

\noindent where $\Delta M_T$ is the mass loss in \msunyr over the time $T$ in years,
$\rm{X_j}$ is the mass fraction of species $j$, and $\rm{X_{j,init}}$ is the initial mass fraction of that species. 

At the point of core collapse we define the core compactness parameter as \citep{oconnor:11}:

\begin{equation}\label{eq:compactness}
    \xi_{M} = \frac{M/\msun}{R(M=2.5\msun)/1000\rm{km}}
\end{equation}

\noindent where $M$ is the enclosed mass (taken as 2.5\msun) and  $R$ is the radius at this
mass coordinate in \rm{km}. As there are multiple criteria possible for determining the explodability of a model, we also use the criteria of \citet{ertl:15} following the \texttt{w18.0} fit, defining models to explode as those with:

\begin{equation}\label{eq:ertl}
    \mu_4 < 0.283 M_4 \mu_4 + 0.0430 
\end{equation}

\noindent where $M_4$ is the mass coordinate (in solar masses) where the entropy per nucleon S = 4, and $\mu_4$ is the mass gradient at this mass coordinate, defined as:

\begin{equation}\label{eq:mu4}
    \mu_4 = \left.\frac{\Delta m/\msun}{\Delta r/1000\rm{km}}\right\vert_{S=4}
\end{equation}

\noindent where we adopt a $\Delta M=0.3\msun$ and $\Delta r$ is the change in radius at S=4 over $\Delta M$.

\subsection{Core-collapse supernovae}

Our explosion models follow those of \citet{farmer21}.  We first excise the 
the inner region of each star's core, by placing the inner boundary of our model
at the point where the entropy per baryon $\rm{S}=4$ \citep{brown13}. We then use a thermal bomb, placed at the inner boundary
of the model to inject energy into the star \citep{aufderheide91,sawada19}.
We do not allow the model to collapse further inwards after the excising the core, which can alter the final nucleosynthetic signal \citep{imasheva22}.
In our default model, energy is then injected
over the inner $0.01\msun$ and over 0.0045 seconds. We inject
sufficient energy to bring the total energy of the star (the sum of the kinetic plus thermal energy minus the gravitational binding energy) to $10^{51} \ergs$.  While there is some evidence for choosing different explosion parameters between stripped envelope supernovae and non-stripped supernovae \citep{saito22}, we chose to use a constant set of parameters to minimise the parameter space explored. 
In Section \ref{sec:sn_uncert} we show the uncertainty introduced into the final yields due to differences in the explosion parameters.

The energy we inject into the star is sufficient to drive a hydrodynamic shock 
that travels outwards from the center of our model to the surface. We stop our models once
the shock reaches $0.1\msun$ below the surface. In \cite{farmer21} we found that the final total energy of a star would be 5--20\% higher than what was injected into the star during the explosion. This
was due to the change in the nuclear networks (between the end of the stellar evolution and the start of the supernovae) introducing new isotopes that were not in
nuclear statistical equilibrium (NSE)\footnote{\MESA{} does not assume NSE, instead it uses the same nuclear network integration that is used at lower temperatures \citep{paxton:15}}. In this work, as we have used the same nuclear network for
the entire stars evolution, the core maintains NSE between the stellar evolution and core collapse phases. The difference in final total energy of the star compared to the total energy after the energy injection phase is now $<1\%$.

\subsection{Stable isotopes}

Many of the isotopes in our 162 isotope nuclear network are short-lived radioactive isotopes and thus
would not be expected to contribute to the Galactic yields. Therefore, we also compute the set of stable isotopes converting each radioactive isotope into its single most probable decay product (Timmes, F., priv.\ com.). This is effectively assuming that we observe the isotopes on a timescale much longer than their half-lives, and that each isotope has only one decay channel\footnote{This assumption is reasonable for most isotopes in our network except for \chlorine[36], \scandium[46], \manganese[54], \cobalt[58], \copper[64] where additional decay channels can play a role.}.

\section{Total Ejecta}\label{sec:yields}

\begin{figure*}[ht]
     \centering
     \subfigure[All isotopes]{\label{fig:imf_all}\includegraphics[width=0.49\linewidth]{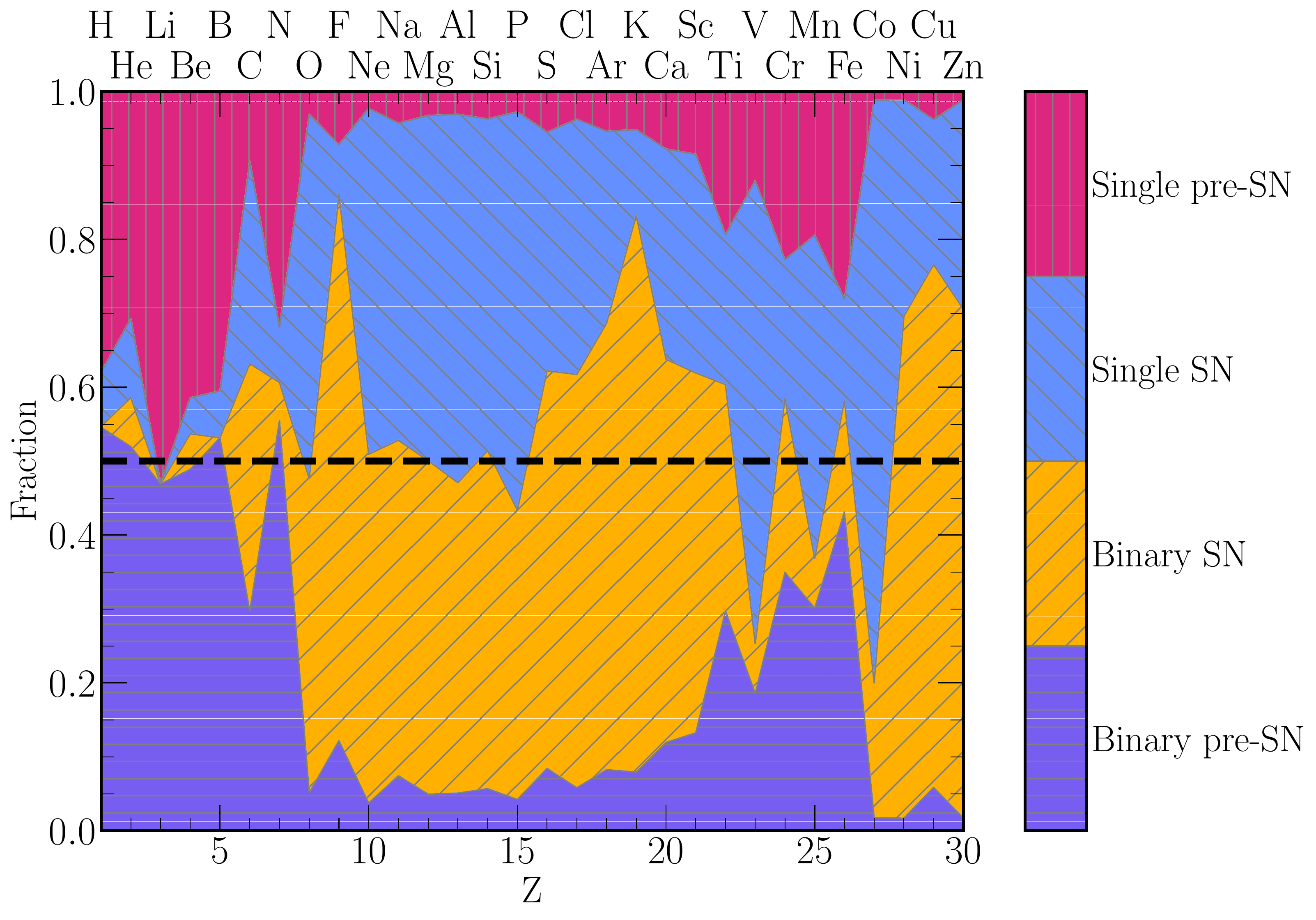}}
     \subfigure[Stable isotopes]{\label{fig:imf_stable}\includegraphics[width=0.49\linewidth]{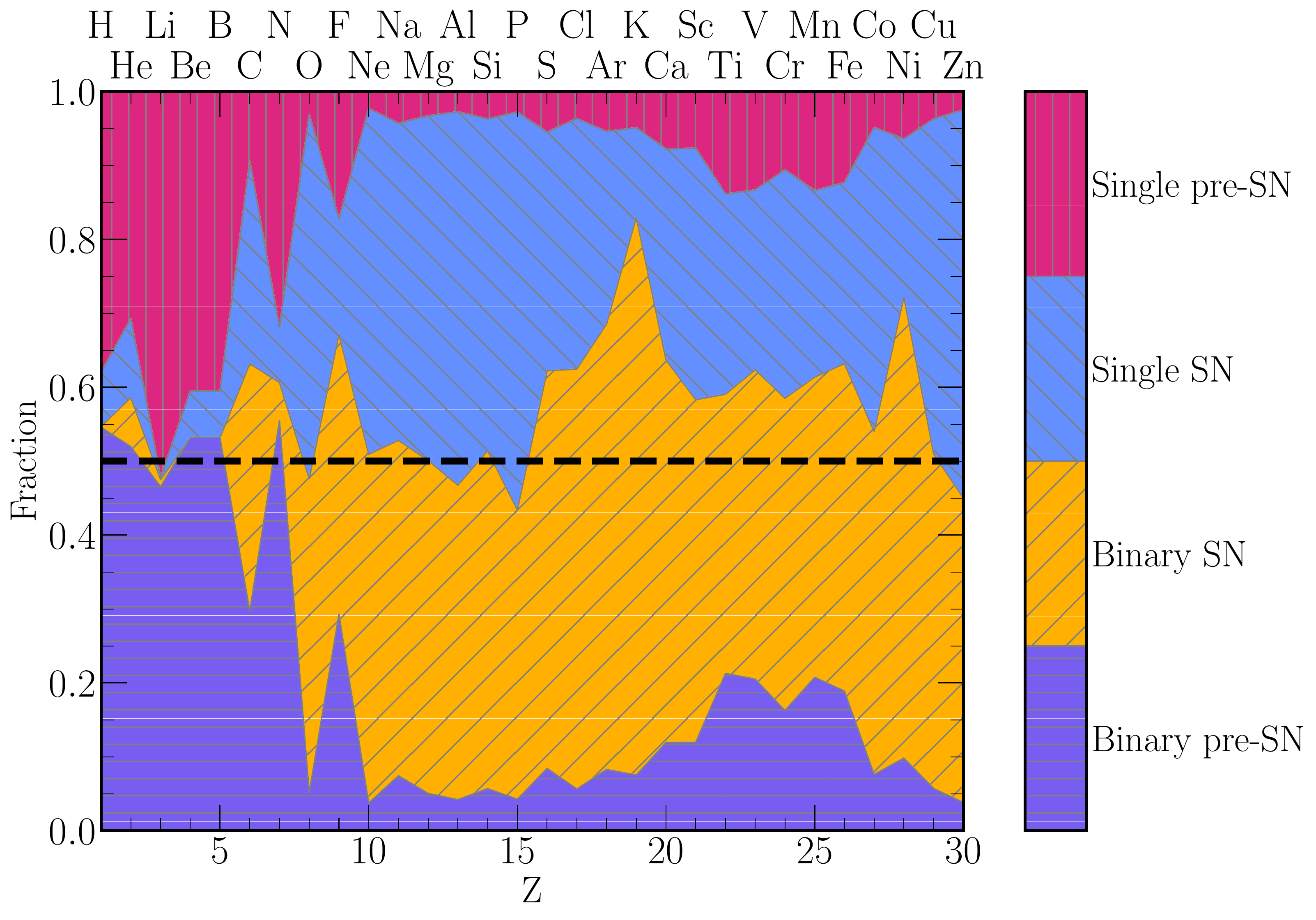}}
        \caption{ Fraction of the IMF-weighted ejecta (not yields) per element as a function of the different formation channels. Pre-SN denotes all pre-supernova mass loss (wind mass loss and RLOF),
  while SN denotes the ejecta from successful SN (using the \citet{ertl:15} criterion). Panel (a) shows the contribution from all isotopes, while panel (b) shows only the stable isotopes after decaying all radioactive isotopes to their most likely decay product. Vertical (horizontal) hatching shows the single (binary) star pre-supernovae mass loss while reverse (forward) diagonal hatching shows the single (binary) supernovae mass loss. }
        \label{fig:imf_fraction}
\end{figure*}

We define the fraction of each isotope $F_i$ which is ejected by either single or binary stars in different mass-loss processes as:

\begin{equation}\label{eq:imf_yield}
    F_i = \frac{\int_{11}^{45} \left( E_{pre,i} + f_{SN}E_{post,i}\right) m^{\alpha} dm}{\sum_{b,s} \int_{11}^{45} \left( E_{pre,i} + f_{SN}E_{post,i}\right) m^{\alpha} dm}
\end{equation}

\noindent where $E_{pre,i}$ ($E_{post,i}$) is the ejected mass of isotope $i$, from the pre-SN mass loss (post-SN mass loss) from either the binary-stripped ($b$) or single stars ($s$). This is then integrated over the mass range $m$ (in solar masses) and weighted by the initial mass function (IMF), with an assumed power law $\alpha=-2.3$ \citep{salpeter:54,schneider18}. The factor $f_{SN}$ is a term that is set to control whether or not we include the contribution from a star's post-SN mass loss, depending on whether the envelope is ejected ($f_{SN}=1$) or not ($f_{SN}=0$). This is then normalized by summing over the IMF-weighted contributions from all mass loss processes for both single and binary stars.
For the binary-stripped models which did not evolve to CC, we interpolate their wind yields over the failed models, and assume that they would not successfully eject their envelopes during core collapse.

Figure \ref{fig:imf_fraction} shows the results of Equation \ref{eq:imf_yield}, which breaks down whether each element was ejected during wind mass loss or RLOF assuming the secondary star was able to later eject the material unprocessed (pre-SN), or in the supernova ejecta (SN). For the supernovae ejecta we assume the explosion model given by equation \ref{eq:ertl} of \citet{ertl:15} (See section \ref{sec:explode}) to set the filter term $f_{SN}$. Panel (a) shows all elements while panel (b) shows the result of decaying all unstable isotopes to their most probable stable product. This assumes that the ratio of single stars to binary-stripped stars is approximately equal \citep{sana:12}.

We draw several broad conclusions from Figure \ref{fig:imf_fraction}. Firstly, massive binaries in general produce larger fractions of most elements compared to massive single stars (50--60\% of the total yield). Most light elements ($Z<8$), lighter than oxygen, are ejected during wind mass loss (or RLOF), while newly-synthesized heavier elements
unsurprisingly require a supernova to be ejected. Binary-stripped stars are also able to eject a significant fraction
of unstable isotopes with $20<Z<27$ (calcium to cobalt) in their pre-SN mass loss (See section \ref{sec:winds}).
Stable elements in the $20<Z<27$ (calcium to cobalt) range, in pre-SN mass loss, are produced in their helium cores via neutron captures. Thus stripped binaries are able to eject more of those isotopes than single stars, as they expose their helium cores earlier, allowing for a greater total amount of mass to be ejected from their helium cores. The same effect is responsible for the increased \carbon[12] production in binary-stripped stellar winds \citep{farmer21}.

The large spike at Z=9 (fluorine) for the binary SN ejecta in Figure \ref{fig:imf_all} is due to the production of the short lived isotope \fluorine[18] which decays to \oxygen[18]. The source of stable fluorine is still uncertain \citep{renda04,lugaro08,franco21}, but Figure \ref{fig:imf_stable} indicates that massive binaries can make more overall than massive single stars. Though the final yield will depend on neutrino spallation of \neon[20] in the supernovae shock, which we do not follow \citep{woosley:88,woosley:02}. This increased production of fluorine is due to two factors, firstly the wind yields are higher in binary-stripped stars, due to positive yields only occurring in stars that expose their C/O cores during core helium burning \citep{meynet00b}. Secondly the supernovae yields peak for the lowest-mass binary-stripped models, which are then favoured by the IMF. 

At core collapse \fluorine[19] is produced via the chain $\nitrogen[14]\left(\alpha,\gamma \right)\fluorine[18]\left(\beta^{+}\right) \oxygen[18]\left(p,\alpha \right)\nitrogen[15]\left(\alpha,\gamma\right)\fluorine[19]$ \citep{goriely90,limongi18}, at the base of the helium shell. In the binary-stripped models this shell is slightly cooler ($\log (T/K) \approx 8.3$) than in single stars ($\log (T/K) \approx 8.5$) at core collapse.  This leads to an increased destruction of \oxygen[18] in the single star models, which limits the production of \fluorine[19].

Figure \ref{fig:imf_fraction} shows that potassium is the element most affected by binary physics, with most of its contribution coming from the SN ejecta of binary-stripped stars. Potassium abundances have been challenging to fit in previous models of chemical enrichment \citep{kobayashi20}, which suggests the binary-stripped stars may be the key to its production. This peak is predominantly from \potassium[39] produced by our low mass binary-stripped stars during their pre-supernovae evolution. The potassium is produced just outside what will become the compact object, inside the silicon rich-layers. The amount of \potassium[39] produced is correlated with the size of this layer and thus may depend sensitively on the uncertain properties of convection during Si/Fe burning. 
 
Around $Z\approx25$ (manganese), single stars produce more odd-Z nuclei in their supernovae while even-Z nuclei are produced equally by single and binary stars. This is due to vanadium being produced in the single stars that are still red super giants (RSG) at core collapse (these have a low initial masses and are thus highly weighted by the IMF). This vanadium then forms a seed for alpha captures that then produce manganese and cobalt. Stars that have lost their H-envelopes (either stars stripped in binaries or high-mass single stars) produce much less vanadium during core collapse, thus there are less seed nuclei present for this odd-even pattern. This pattern disappears however when considering only the stable decay products, therefore a detectable signal will depend heavily on the timing of the observations relative to the explosion.  

In the online Zenodo material we provide the yields for all isotopes from all mass-loss processes, as well as diagnostic information about the stellar models. See Appendix \ref{sec:data_table} for an example of the data provided.

\section{Pre-supernova evolution}\label{sec:res_winds}

During the pre-supernova evolution of our models the chemical yield is affected by both the wind mass loss and the mass lost during RLOF (for binaries). This difference in mass-loss processes has two effects on the yields, firstly it changes the total mass loss and the timing when that mass is lost, secondly the mass loss has an effect on the core structure of the star leading to changes in its composition and core mass \citep{kippenhahn:67,habets86,laplace21}. In general the greater the fraction of the star's mass that is lost, and the earlier that it is lost in the star's life, the lower the final core mass becomes.

\subsection{Composition of stellar winds}\label{sec:winds}

Figure \ref{fig:imf_fraction} shows that the pre-supernovae mass loss contributes almost the entirety of the
chemical yields for low-mass elements. For higher-mass elements, massive binary-stripped stars are able to eject $\approx20\%$ of their total ejecta in iron-group elements, while for massive single stars this is closer to $\approx10\%$. This occurs due to the increased mass loss in the binary-stripped stars exposing the He-core during core helium burning. This material is enriched in neutron capture isotopes
(see also the much larger contribution of the binary pre-SN mass loss in Figure \ref{fig:imf_fraction}a).
These neutrons come primarily from $\neon[22]\left(\alpha,n\right)\magnesium[25]$ \citep{prantzos87}.
See Appendix \ref{sec:winds_eta} for a discussion of the sensitivity of the isotopic yields due to the wind-scaling factor, and hence total wind mass loss.

We can divide the yields into 3 broad categories: 1) those that change approximately linearly with mass lost (e.g.\ \hydrogen, \helium) either positively or negatively; 2) those that are small or negative until the star exposes its helium core (e.g.\ \carbon{} or \fluorine[19]) before increasing; 3) those that fall into neither of those categories, which we discuss below.

\subsubsection{\oxygen[17]}

For massive single stars \oxygen[17] has positive yields only for stars with $\mint<31\msun$, with the yield increasing as the initial mass decreases. The peak formation is at $\mint=18\msun$ before the yield then decreases at even lower masses. For massive binary stars the yields are positive for masses below $\mint\lessapprox 25\msun$, with no turnover in the yields. The stripped binaries have larger yields than single stars below $\mint<16\msun$.

\subsubsection{\magnesium[24]}

Both our stripped binaries and single stars show positive yields for \magnesium[24]. However,
for single stars the yields peak at $\mint=36\msun$ (when the star exposes its helium core), before decreasing at higher masses. In contrast, for the binaries the yield always increases as the initial mass increases. 

\subsubsection{\vanadium[51]}

We find that \vanadium[51] is the only isotope in our nuclear network that has a positive
wind yield from binary-stripped massive stars while having a negative wind yield from single stars. While this would make wind mass loss from stripped-binaries an candidate source of \vanadium[51] we find that the total yield of \vanadium[51] is dominated by the
core-collapse yields from single stars. Electron captures onto vanadium can play a key role in setting the final electron fraction of the core before collapse \citep{aufderheide94,heger01}.

\subsection{Composition of mass lost during RLOF}\label{sec:rlof}

\citet{farmer21} showed that the mass transferred during Case B RLOF stripping is not enriched in \carbon[12] compared to the initial abundances, however the structural changes caused by the extra mass loss would affect the star's evolution, and thus the final \carbon[12] yield. This was due to the fact that \carbon{} was primarily produced during core helium burning which occurred \textit{after} mass transfer finished. 

However, there are also a number of isotopes that are produced in hydrogen envelopes and thus can be ejected \textit{during} the mass transfer process. These are generally light elements that are either produced in the low temperature environment of the hydrogen envelope or are able to be mixed out of the hydrogen core/shell into the envelope and then ejected.

Figure \ref{fig:imf_stable} shows similar total ejecta from single and binary-stripped massive stars. However, the net contribution to the Universe will be reduced in the binary-stripped stars as compared to single stars, as their contribution to the composition of the Universe will depend sensitively on how conservative mass transfer is and how the secondary reprocesses and re-ejects any accreted material. We assume for the calculation of the RLOF phase  that all the material we eject from the donor accretes onto the accretor (fully conservative). However, we track these isotopes separately and can infer what the composition might be if some fraction of this accreted material is ejected during RLOF or is re-ejected later by the accretor without further nuclear processing as a wind. We will note when we include the contribution from RLOF (or not). Overall, for elements Phosphorus (Z=15) and heavier, the absolute yields from RLOF are $<1\%$ of the absolute wind yields. While for lighter elements the absolute RLOF yields can be comparable to the absolute wind yields.

\begin{figure}[ht]
  \centering
  \includegraphics[width=1.0\linewidth]{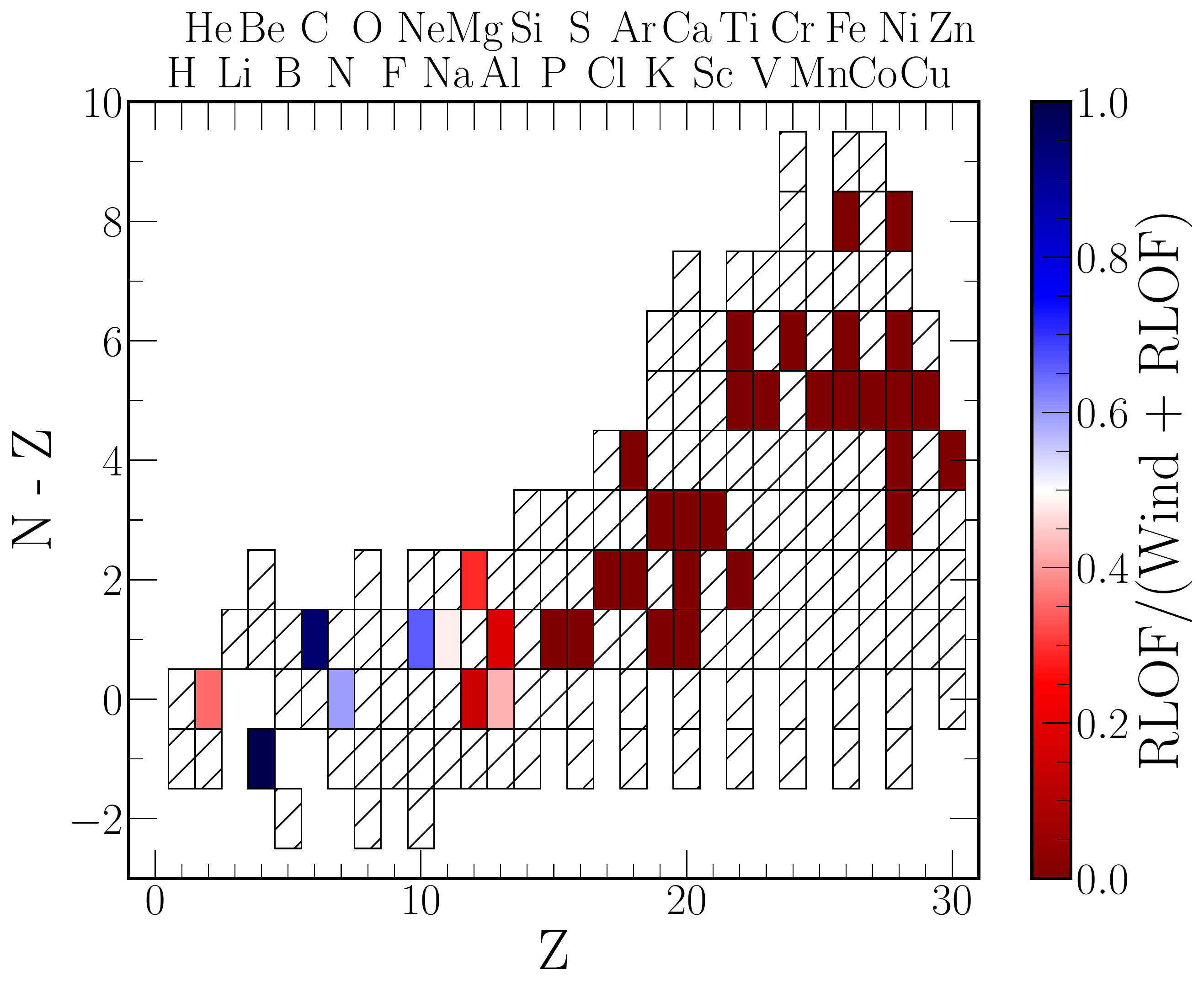}
  \caption{ The IMF weighted relative yield fraction of ejected isotopes from RLOF compared to wind mass loss in binary systems. The x-axis is the proton
  number, Z, while the y axis is the number of neutrons (N) minus the number of protons (Z). Hatched squares indicate that one or both ejection processes (RLOF or winds) have negative yields or the yields from
  either RLOF or wind mass loss are below $10^{-15}\msun$.}
  \label{fig:rlof}
\end{figure}

Figure \ref{fig:rlof} shows the relative contribution to the IMF-weighted yield from pre-SN mass loss, due to RLOF and stellar winds.
Red regions signify that the isotope is predominately ejected in stellar winds, while blue signifies that the isotope is predominately ejected during RLOF. 
The blue regions show that \carbon[13], \nitrogen[14], and \neon[21] are primarily ejected during RLOF (we ignore \beryllium[7] from this discussion, see section \ref{sec:ignore_neut}).

As the hydrogen-burning core recedes it will leave behind CNO processed material, allowing the envelope to be enriched in \nitrogen[14] which is then ejected during RLOF. The NeNa-cycle is able to convert hydrogen to helium (similar to the CNO cycle) via neon, sodium, and magnesium. 
Above temperatures of $T>3.5\times10^7$K \neon[21] is destroyed via $\neon[21]\left(p,\gamma\right)\sodium[22]\left(\beta^+\right)\neon[22]$, while below this temperature \neon[21] is produced via $\neon[20]\left(p,\gamma\right)\sodium[21]\left(\beta^+\right)\neon[21]$
\citep{wiescher86,jose99}. The amount of \neon[21] is maximised when there is H burning in the temperature range $3 \leq T/10^7 \rm{K} \leq 3.5$ \citep{arnould99b}. Thus significant amounts of \neon[21] can only be ejected if the material is burnt in the H-shell and mixed outwards before the H-shell can further process the material.

\carbon[13] can provide a significant fraction of the neutron flux at stellar densities and temperatures, via the reaction $\carbon[13]\left(\alpha,n\right)\oxygen[16]$ \citep{busso99},
thus its production (and destruction) plays a key role in the s-process \citep{gallino98}.
The ratio of \carbon[12]/\carbon[13] has also been used as a a probe of galactic chemical evolution \citep{milam05}.

\begin{deluxetable}{ccc}
\tablecaption{The ratio of the IMF integrated \carbon[12]/\carbon[13] ejected during different mass loss processes.}\label{tab:c12c13}
\tablehead{
 & \colhead{Single} & \colhead{Binary} \\
}
\startdata
Winds & 50 & 802 \\
RLOF & - & 17 \\
\hline
pre-SN total & 50 & 123 \\
post-SN total & 360 & 321 \\
\hline
Total including RLOF &  154 & 193 \\
Total not including RLOF &  154 & 446 \\
\enddata
\end{deluxetable}

Table \ref{tab:c12c13} shows the IMF weighted  \carbon[12]/\carbon[13] ratio from the different mass loss processes. The totals show the  \carbon[12]/\carbon[13] ratio depending on whether the mass donated by the primary is ejected from the system (with RLOF) or is assumed to be accreted onto the accretor and mixed inwards (without RLOF), thus not being ejected into the Universe. This assumes the accretor does not re-eject the material at a later time. Table \ref{tab:c12c13} therefore indicates that the contribution to the Galactic \carbon[12]/\carbon[13] from massive stars will depend sensitivity on how conservative mass transfer is, thus how much mass is accreted, and whether the accretor re-ejects the material or it is mixed into the core.  Thus the \carbon[12]/\carbon[13] ratio in regions where the dominant enrichment is from massive stars could potentially be used as a tracer of how conservative mass transfer is, as the ejection of \carbon[13] occurs primarily during the mass transfer process, while \carbon[12] is primarily ejected from WR wind mass loss. 

The Solar system has \carbon[12]/\carbon[13]$\approx 90$ \citep{romano19,Botelho20}, which is lower than the values in Table \ref{tab:c12c13}, thus the primary producer of \carbon[13] is likely lower-mass stars, for instance AGB and TP-AGB stars which can have \carbon[12]/\carbon[13] values  between $\sim10$--$180$ depending on the initial mass \citep{karakas16}.

\subsection{Core composition}

\begin{figure}[ht]
  \centering
  \includegraphics[width=1.0\linewidth]{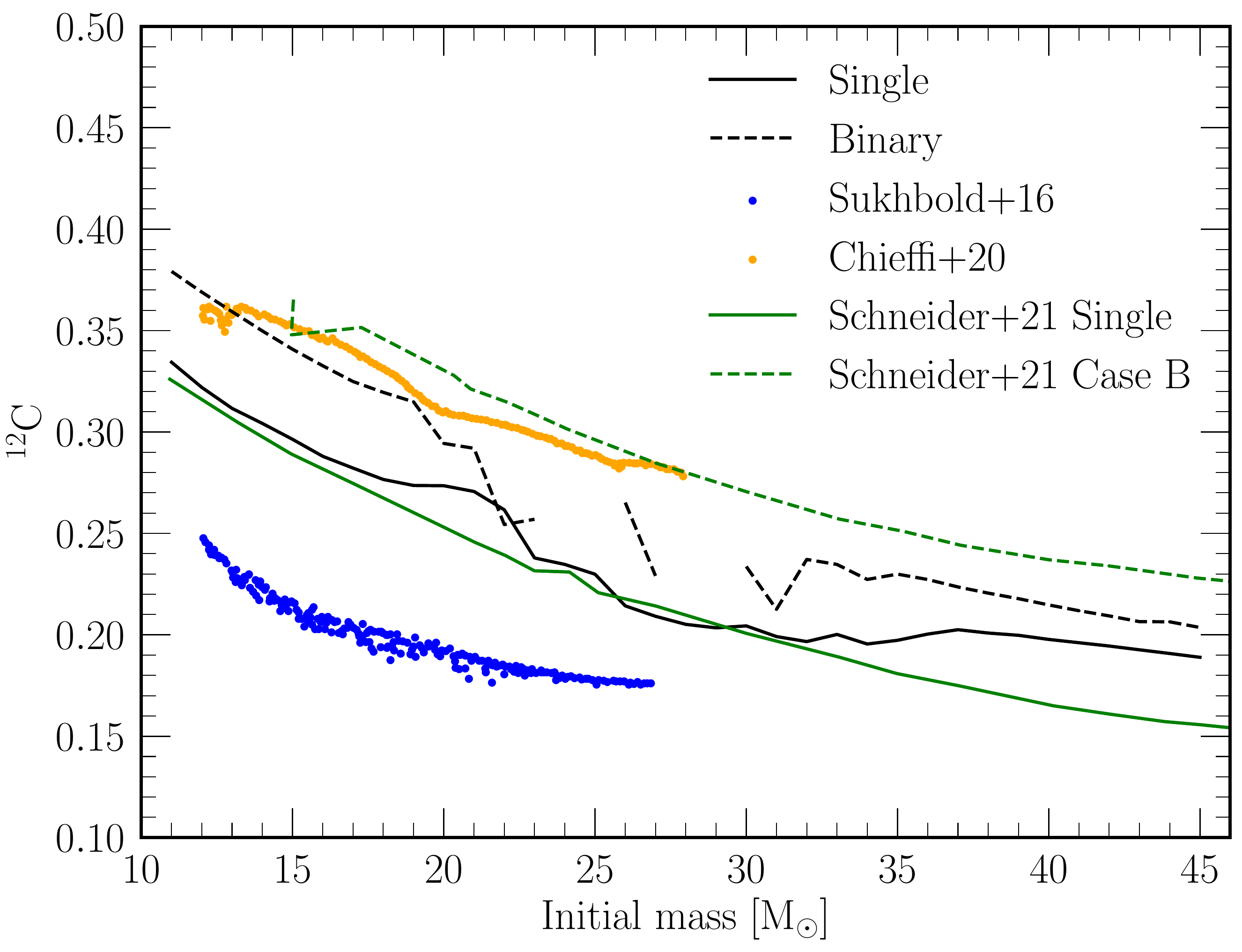}
  \caption{Comparison of \carbon{} mass fractions at the end of core helium burning. Black solid line: Single stars, Black dashed line: Binary-stripped stars, Blue: \citet{sukhbold16}, Orange: \citet{chieffi20}, Green solid line: Single star models from \citet{schneider21}, Green dashed line: Case B binary models from \citet{schneider21}.}
  \label{fig:c12_comp}
\end{figure}

The final fate of a star depends on many quantities, however the core mass and the central \carbon[12] fraction at the end of core helium burning are two of the most important \citep{chieffi20,patton20}. Figure \ref{fig:c12_comp} shows a comparison of the core \carbon[] fraction at the end of core helium depletion for various solar metallicity models. In general we can see that as the initial mass increases, the core \carbon[12]{} fraction decreases. It is set by the combination of the $3\alpha$ rate and the \crate{} reaction \citep{sukhbold20}. For our single star models, the \carbon[12]{} fraction agrees well with \citet{schneider21}, which is unsurprising as \citet{schneider21} also used \MESA{} with similar physics choices, although there is a divergence at high initial mass, which may be due to our different choices of chemical mixing. \citet{schneider21} also uses a similar wind prescription as this work, but with a different dependence on the metallicity. 

Our binary-stripped models also closely match  the values of \citet{schneider21} case-B binaries, with differences likely due to the different method of removing the envelope, as \citet{schneider21} takes the material off with an ad-hoc prescription, while we follow the binary mass transfer. Our models show a variability in the \carbon[12]{} trend, not seen in \citet{schneider21}, in the 20--30\msun{} mass range. This is where both the MLT++ approximation is important and the envelope is not significantly removed by winds before mass transfer begins. Thus the envelopes properties depend sensitively on the ad-hoc MLT++ prescription, which
propagates as an uncertainty in how wind mass loss occurs, and how efficient RLOF is at removing the envelope. How mass is lost from a star has been shown to significantly impact the internal structure of massive stars \citep{renzo:17}.

Our binary-stripped models also agree well with the \citet{chieffi20} single star models, for $\mint<20\msun$. Both \citet{chieffi20} and this work use NACRE \citep{angulo99} for the 3$\alpha$ rate. For the \crate{} rate
\citet{chieffi20} uses \citet{kunz02} while we use \citet{angulo99} rate. Therefore differences likely stem from different choices in the nuclear physics. Thus binary stripping has a similar effect on the core abundance as variations in the \crate{} reaction rate.

As shown extensively in \citet{laplace21} the effect of the extra mass loss due to RLOF causes structural changes to the cores of massive stars. These changes cause the cores to become smaller (in mass) and to have extended chemical gradients. This leads to differences in the intensity and timing of nuclear shell burning episodes, which compounds the differences between single and binary-stripped stars. While \citet{farmer21} has shown how the binary interactions affect the final \carbon[12] yield.

Details of key stellar properties can be found in Appendix \ref{sec:appen_evol} and in Tables \ref{tab:evol_single} and \ref{tab:evol_bin}. These include the final total and core mass, total mass lost to different mass loss processes, and final masses of H, He, and C.

\section{Supernovae}\label{sec:res_sn}

The nucleosynthetic yields from core-collapse supernovae are the result of (i) whether or not the star can successfully explode and produce ejecta and (ii) the pre-collapse structure and composition.

\begin{figure}[ht]
  \centering
  \includegraphics[width=1.0\linewidth]{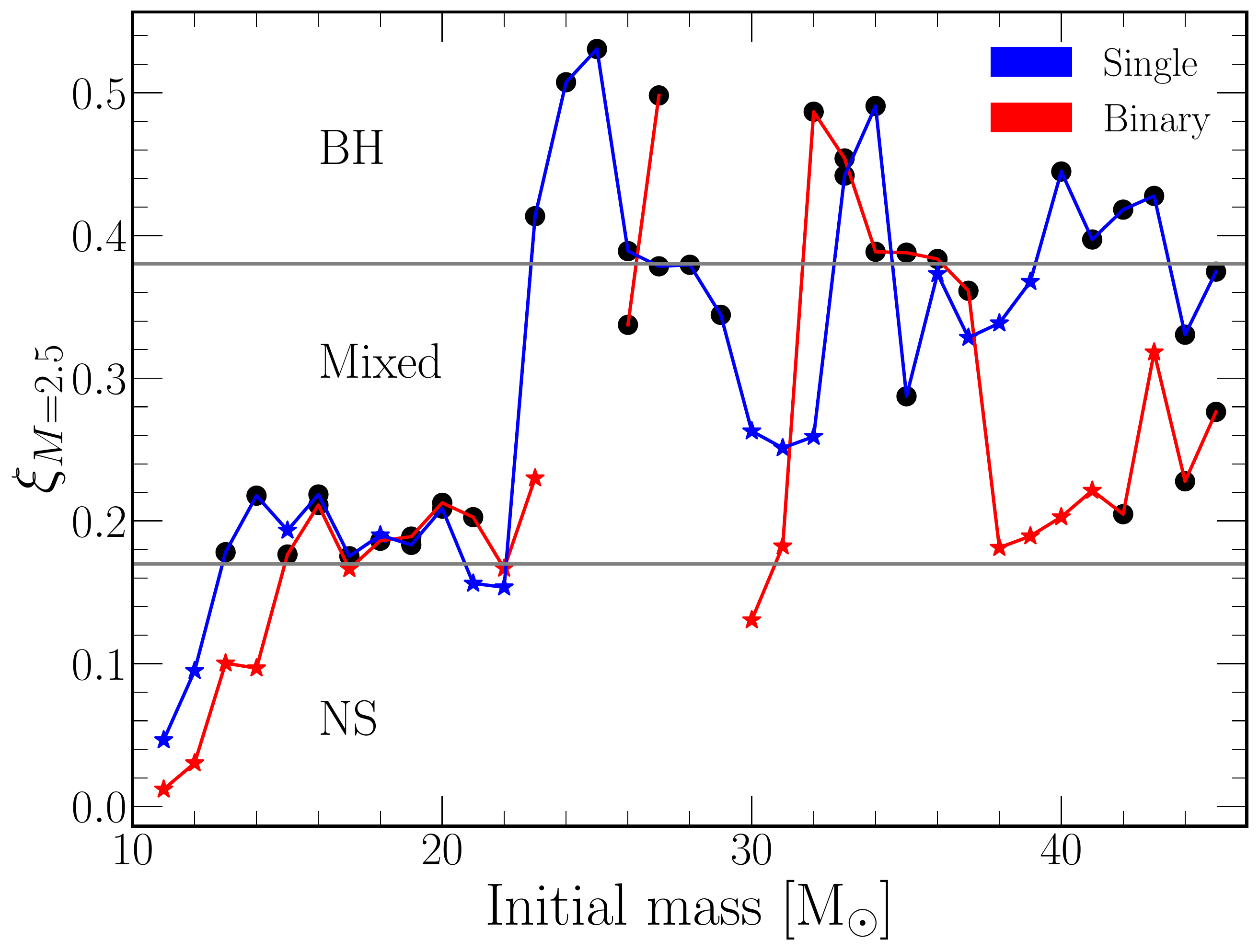}
  \caption{The compactness parameter as a function of the initial stellar mass for massive single stars (blue) and binary stars (red). The grey lines mark
  $\xi_{M=2.5}=0.16$ and $\xi_{M=2.5}=0.36$ (with $dm$ averaged over 0.3\msun) as the 
  approximate boundaries between all models undergoing a successful explosion ($\xi_{M=2.5} \leq 0.16$), no model undergoing a successful explosion ($\xi_{M=2.5} \geq 0.36$), and an 
  intermediate mixed state.
  Star symbols mark models that would successfully explode under the
  two-component fit of \citet{ertl:15} calibrated against the model \texttt{w18.0}, while black circles denote models that would not be expected to produce successful explosions (and thus
  ejecta).
  }
  \label{fig:explodes}
\end{figure}

\subsection{Explodability}\label{sec:explode}

We first determine whether a star will successfully explode and eject its envelope, or not. Figure \ref{fig:explodes} shows two measures of a star's explodability: its compactness \citep{oconnor:11} (y-axis) and the two component fit of \citet{ertl:15} (symbols), as a function of the initial mass. As shown in many other works \citep[e.g.][]{sukhbold16,farmer16,renzo:17,sukhbold20,patton20}, the evolution to core collapse and the star's final structure is sensitive to the choice of physics, treatment of chemical mixing boundaries \citep{chieffi20}, and numerical choices \citep{farmer16} made by the software instrument used. In Figure \ref{fig:explodes} we define three regions based on the two component fit of \citet{ertl:15}\footnote{See also \citet{boccioli22} for a potential new criterion for determining explodability}. One region contains models that only make NSs ($\xi_{M=2.5} \leq 0.16$), another contains those that only make
BHs ($\xi_{M=2.5} \geq 0.36$), and there is an intermediate region in which we consider the outcomes to be less certain.  The boundaries of these regions are almost the same as the boundaries found by \citet{ugliano:12} ($\xi_{M=2.5} \leq 0.15$ for NS and $\xi_{M=2.5} \geq 0.35$ for BH). But the presence of the mixed region shows why the compactness, on its own, is not a good measure for determining the explosion outcome.

In Figure \ref{fig:explodes} we can find several regions that are expected to undergo successful explosions, with only slight movement of the boundaries between single and binary stars. Between $11 < \mint/\msun < 21$ and $31 < \mint/\msun < 36$ stars end their lives with similar compactness parameters (and with similar explosion success metrics). In contrast, between $21 < \mint/\msun < 31$ (though there are a number of missing models here) and $ \mint > 38\msun $ the final states of the stars are very different between single and binary-stripped stars, leading to differences both in the explodability and their final nucleosynthetic yields. 

The structure seen in Figure \ref{fig:explodes} is related to the transition in how different fuels ignite in the core \citep{sukhbold:14,laplace21}. The peak at $\mint\approx25\msun$ is when carbon changes from central to off-center burning while the peak at $\mint\approx34\msun$ is coincident with neon going from central to off-center burning \citep{schneider21}. The shift in the location of the peaks between binary-stripped and single stars is due to the binary-stripped stars having smaller core masses, due to the core receding when RLOF removed the envelope \citep{farmer21}.

We include the fallback prescription of \citet{goldberg19}. Here zones near the inner boundary are removed that have a total energy, when integrated from the core to the zone, that is negative. This is done by moving the inner boundary
of the model to the point where the material has a positive 
total energy, and thus is unbound from the core. This is repeated at each time step.
See Appendix A of \citet{goldberg19} 
and \cite{paxton:15} for the full details of this fallback treatment.
For the models presented here and with our default explosion parameters, the amount of fallback is negligible ($\Delta \rm{M} \lessapprox 10^{-3}\msun$) even in cases that do not formally explode via the \citet{ertl:15} criterion. As the amount of fallback is negligible in these models, for the choices of explosion physics used here, we simplify our modelling by using an explodability criterion. This lack of fallback is driven by our choice of a fixed $10^{51}\ergs$ explosion, lower energy explosions can have larger fallback masses with the prescription as implemented in \mesa{} \citep{goldberg19}.

\citet{schneider21} found that for case B binaries (where they artificially remove the envelope), NS production occurs for
$\mint \leq 31.5\msun$ and $\mint>34.0\msun$, while for single stars the range is between $\mint \leq 21.5\msun$ and $23.5 \leq \mint/\msun \leq 34.0$. These boundaries can shift if fallback accretion is taken into account \citep{timmes96}. 
We find a somewhat more complex boundary between NS and BH formation for $\mint \leq 22\msun$ and $\mint > 36\msun$. Differences between the boundaries found by \citet{schneider21} and those in this work are likely due to differences in the physical assumptions, such as the choice of convective boundary mixing, timing of mass loss, and the choice of nuclear network. These have been shown to lead to large changes in the final structure \citep{farmer16,renzo:17,farmer21}.

\subsection{SN yields}

\begin{figure*}
     \centering
     \subfigure[Supernovae yield]{\label{fig:comp_yield}\includegraphics[width=0.49\linewidth]{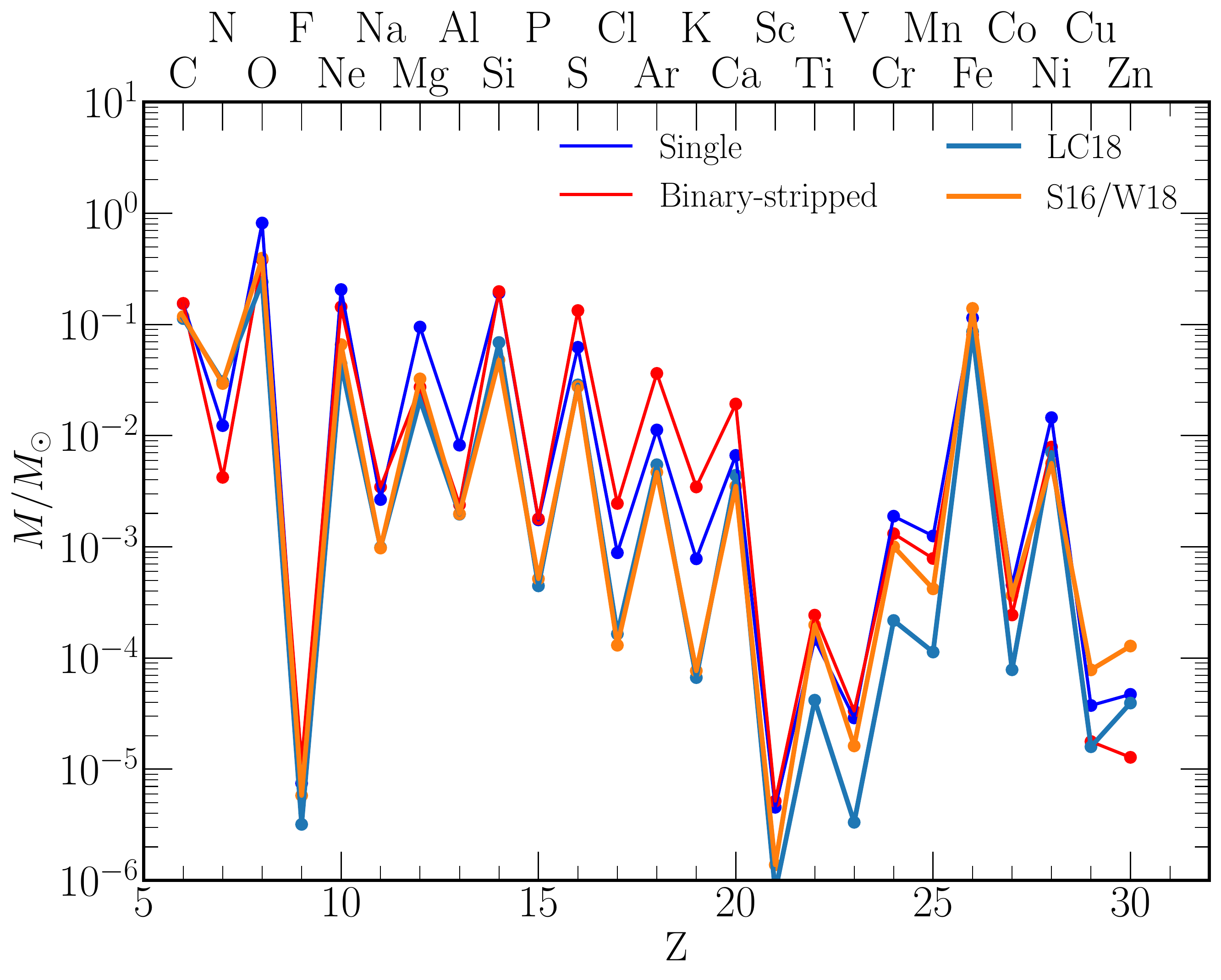}}
     \subfigure[Ratio of yields compared to our single star model]{\label{fig:comp_ratio}\includegraphics[width=0.49\linewidth]{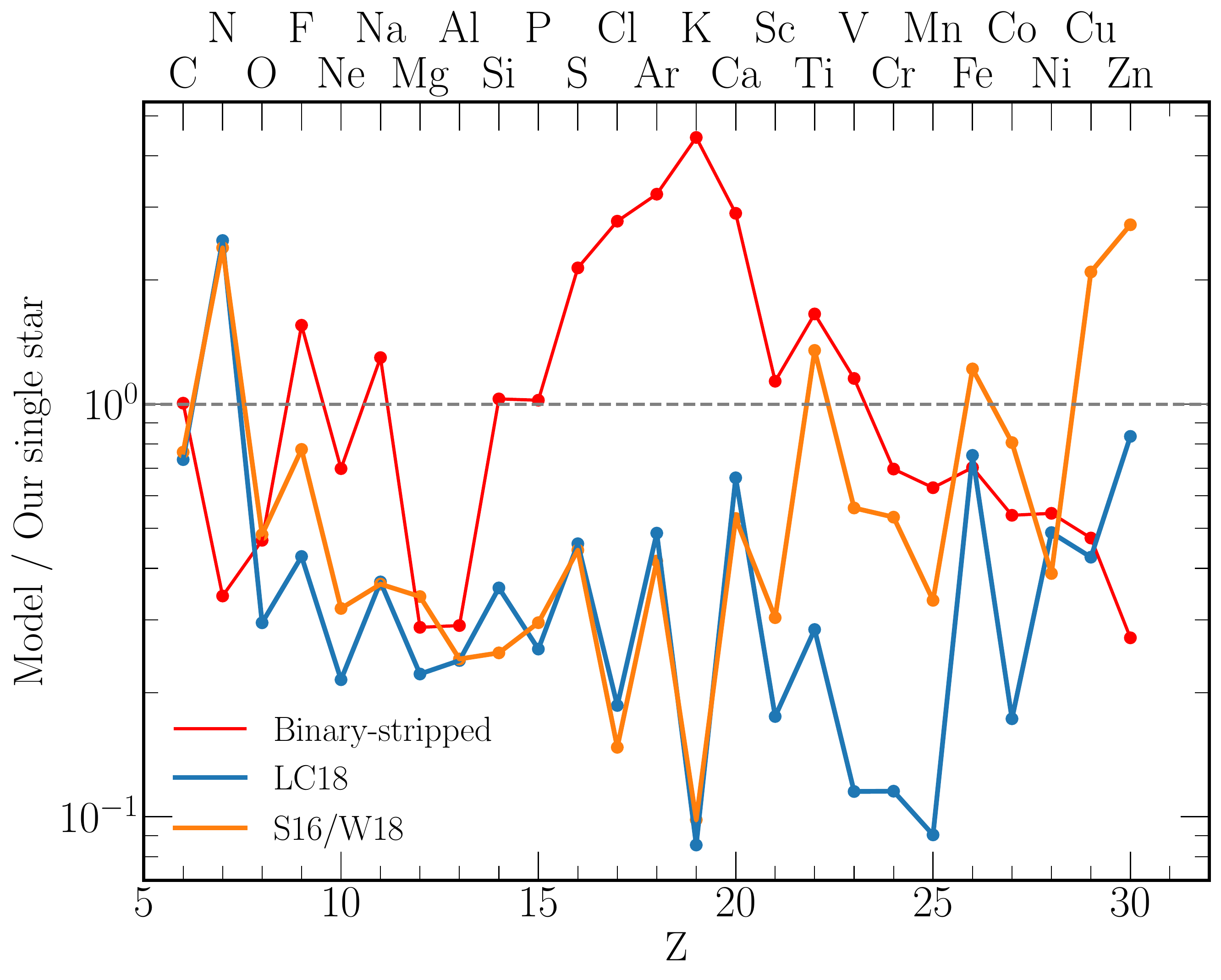}}
        \caption{Panel (a): The stable isotopic yields from core collapse for a 13\msun{} star from the literature; LC18 \citep{limongi18}: light blue, S16/W18 \citep{sukhbold16,woosley:18}: orange) as compared to our $\mint=13\msun{}$ single (dark blue) and $\mint=13\msun{}$ binary-stripped star (red). Panel (b): The ratio of the core collapse
        stable yields from the literature and our $\mint=13\msun{}$ binary-stripped star compared to our $\mint=13\msun{}$ single star.}
        \label{fig:comp}
\end{figure*}

Figure \ref{fig:comp} shows a comparison between our 13\msun{} single\footnote{Formally our 13\msun{} single star does not explode by any of the fits provided in \citet{ertl:15}, however given the limited range of initial masses in \citet{limongi18} that overlap with our models, we chose this initial mass as the binary-stripped model comfortably forms a NS, while both the 15\msun{} models are on the edge between NS and BH, the 20\msun{} models both make BHs, and we have no model for a 25\msun{} stripped-binary. } and binary-stripped stars with yields available in the literature\footnote{We made use of the \texttt{VICE} \citep{johnson20,johnson21,griffith21} software package to provide this information.}. Panel (a) shows the total elemental yields (as a sum of the stable isotopes), while Panel (b) shows the ratio
of the elemental yield for our single star model as compared to the literature values. Our supernova models were exploded with our default explosion physics.

Overall, we over predict (as compared to the literature) elements in the Z=10--21 range by a factor of 5. For higher mass elements we have a good agreement with the \texttt{Kepler} based results \citep{sukhbold16,woosley:18} while diverging to a factor 10 over prediction compared to the results of
\citep{limongi18} which are based on the \texttt{FRANEC} code. Our large disagreement for Zinc is due to the fact that our nuclear network stops at Zinc, thus isotopes will pile up at the edge unable to produce heavier isotopes. 
Other differences are likely due to differences in the stellar micro-physics and physical assumptions. For instance, we showed in \citet{farmer21} that the, currently unconstrained, choice for the amount of overshoot during carbon burning can have a significant impact on the carbon yields, that would then propagate as an uncertainty on the product of carbon burning namely \neon[20], \sodium[23], and \magnesium[24]. This is due to a \carbon[12] yield being sensitive to the size of the pocket of \carbon[] that survives, between the top of the final carbon-burning shell and the helium shell \citep{farmer21}. This pocket, however, can be eroded by convective overshoot from the final convective carbon-burning shells.

Figure \ref{fig:comp}b also shows the comparison between a binary-stripped model and a single star
model, for the same initial mass. Here we can see that the binary-stripped model produces more of the
intermediate-mass elements ($15 < Z < 24$) by up to a factor $\approx4$, while producing slightly
less of the higher mass elements ($Z>24$), as well as producing less magnesium and aluminum (Z=12 and Z=13).

\subsection{Explosive nucleosynthesis}

\begin{figure*}
     \centering
     \subfigure[Single stars]{\label{fig:nuc_exp_sing}\includegraphics[width=0.49\linewidth]{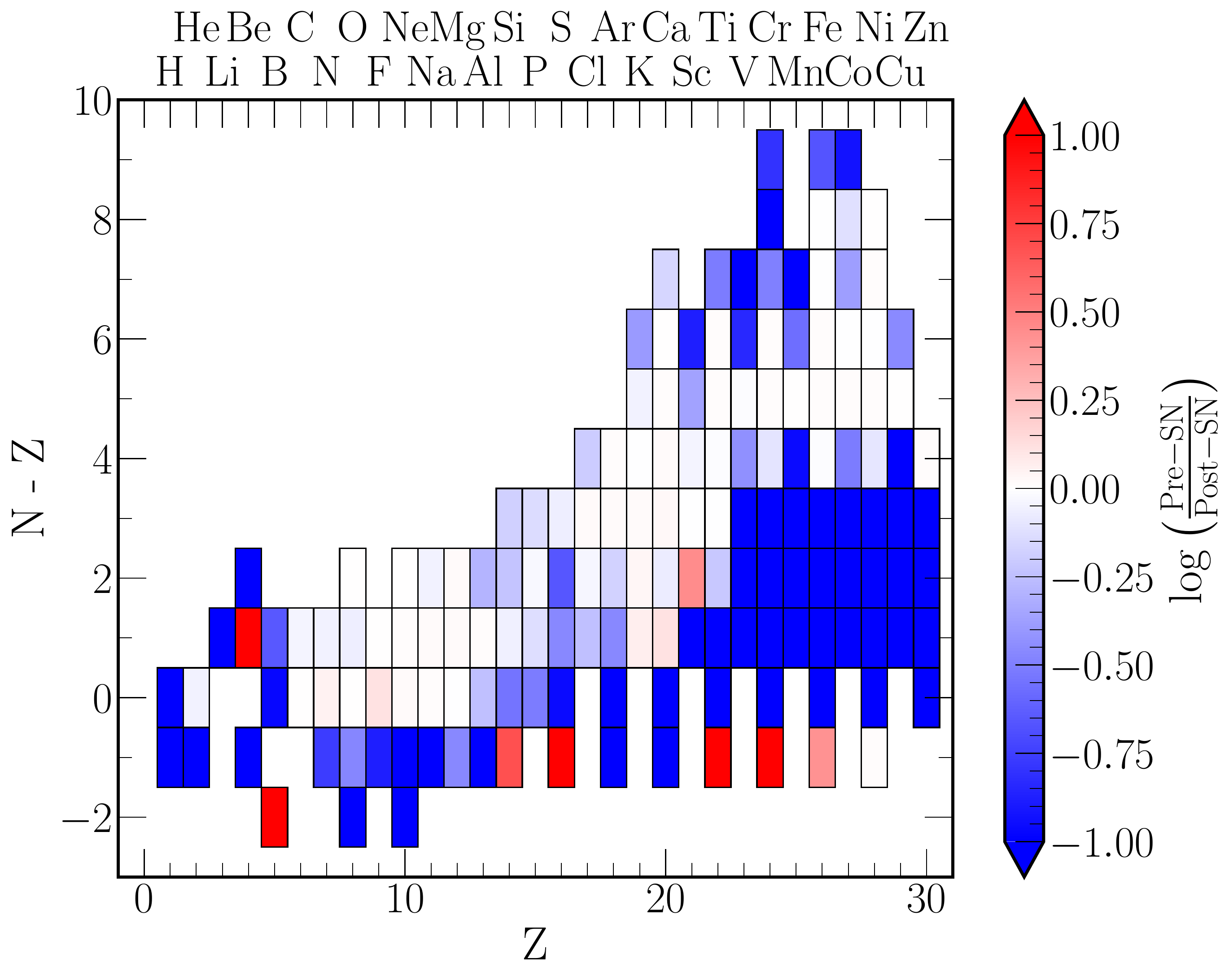}}
     \subfigure[Binary-stripped stars]{\label{fig:nuc_exp_binary}\includegraphics[width=0.49\linewidth]{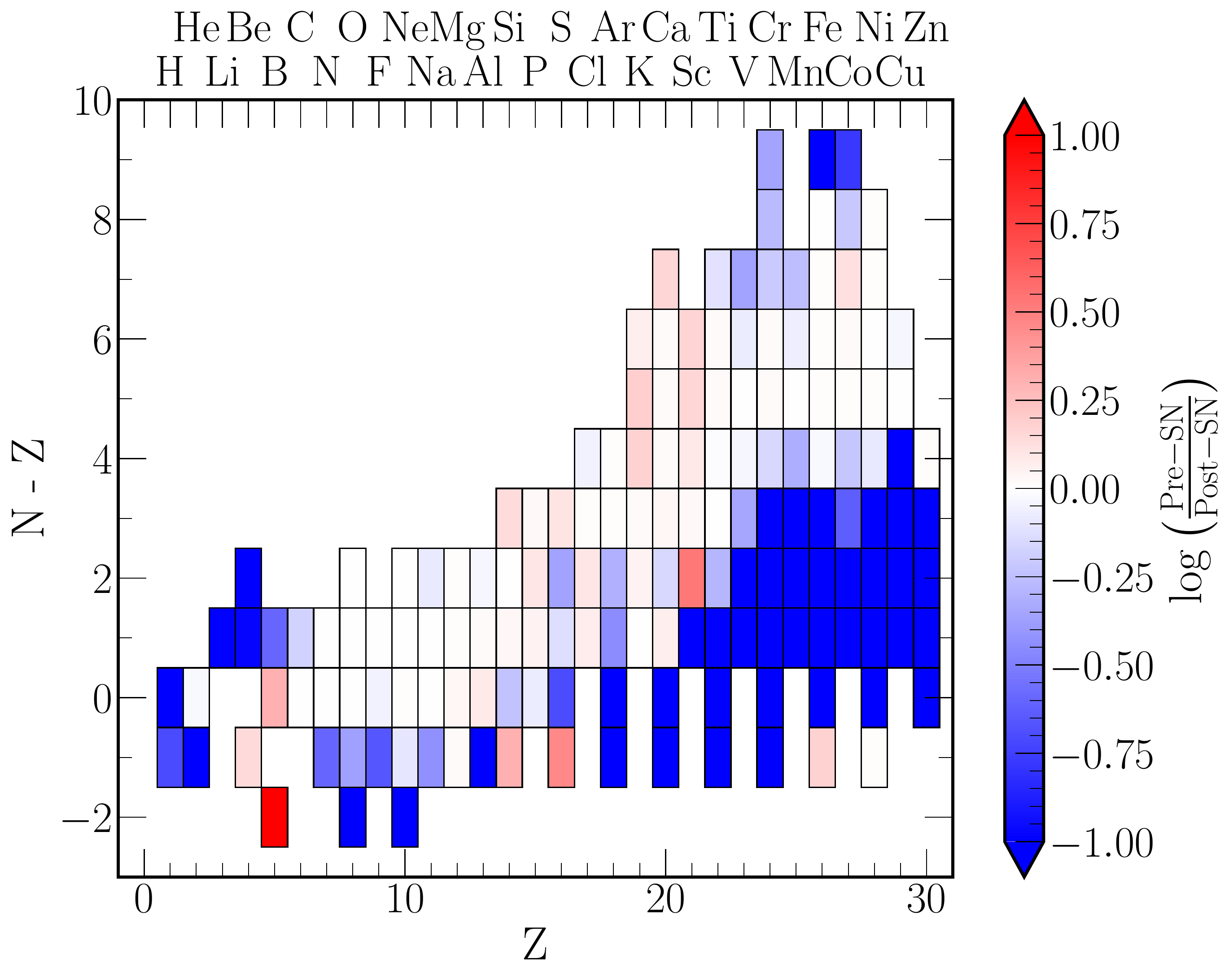}}
        \caption{The logarithmic ratio of the IMF weighted pre-SN composition, outside of the mass cut, to the IMF weighted composition of the models near shock breakout. Blue indicates isotopes produced in the SN, while red indicate isotopes destroyed in the SN. 
        We did not decay any isotopes to their stable versions.
        The colors are truncated to be between -1 and 1.}
        \label{fig:explosive_nuc}
\end{figure*}

Figure \ref{fig:explosive_nuc} shows the relative change in the IMF-weighted material ejected in a SN between its pre-SN composition and its post-SN composition. Blue regions indicate isotopes that are produced in the SN while red regions indicate isotopes that are destroyed in the SN. The color bar has been truncated, to better show the details, however the (blue) material produced can peak at $\approx 10^{15}$ times more (for isotopes in the bottom right corner), and the (red) material destroyed peaks at only $\approx10^4$ times less.

We can see in Figure \ref{fig:explosive_nuc} that most of the region of change occurs for $Z > 20$ and $N-Z < 4$. These are generally the proton-rich isotopes, that have very low abundances outside of the mass cut pre-SN. Thus even a small production during the SN causes a large relative increase in the composition. There is also an increase in the more neutron rich isotopes, typically expected for r-process, in the top right of each panel in Figure \ref{fig:explosive_nuc}. This includes the isotope \iron[60] which is of interest for gamma-ray observations \citep{mahoney82,leising94}. 

Comparing Figures \ref{fig:nuc_exp_sing} to \ref{fig:nuc_exp_binary}, we can see that both show this enhancement in proton-rich isotopes, though the single stars show a greater fraction of material coming from the pre-SN evolution for the most proton-rich isotopes of Ti, Cr, and Fe. For the more neutron-rich isotopes ($N-Z \approx 8$), the single stars show material coming more from the explosive nucleosynthesis during the SN than before. For neutron-rich isotopes of Al--Sc, binary-stripped stars are generally producing this material pre-SN (and outside what will becomes the compact object), while in single stars this occurs during the explosive nucleosynthesis in the SN. 

We reconfirm our finding the \carbon{} is not affected by the supernova shock \citep{farmer21}. We find, with our larger number of isotopes, that \carbon[12]{} is in fact the isotope least affected by the supernova explosion. This is due both to its location near the He-shell, thus little \carbon[12]{} is lost to the formation of the compact object, and when the shock finally reaches the He-shell it has cooled sufficiently to be unable to drive additional \carbon{} burning.

\subsection{SN uncertainties}\label{sec:sn_uncert}

Following the method of \citet{farmer21} we explore the sensitivity of our SN yields to the uncertain explosion physics. We took a 17\msun{}\footnote{Our 17\msun{} single star model does not formally explode given the fits in \citet{ertl:15}, though it does have a similar compactness to the 17\msun{} binary-stripped model.} single and binary-stripped star models and exploded each of them $\approx 100$ times. For each sample we would randomly vary the following four parameters: the injection energy, mass cut, injection mass, and injection time.
As an improvement over \citet{farmer21} we sample all four explosion parameters simultaneously as opposed to doing two grids, which sampled only two parameters at a time. We vary the injection energy between $0.5$--$5.0 \times 10^{51} \ergs$, the mass cut between $1.0$--$2.0\msun$, the injection mass $10^{-2}$--$5\times 10^{-1}\msun$, and the injection time $10^{-3}$--1.0s. We assume a thermal bomb for our explosions, however a piston driven explosion may produce different yields for the same input energy \citep{young07}.

\begin{figure*}
     \centering
     \subfigure[Single stars]{\label{fig:exp_sing}\includegraphics[width=0.49\linewidth]{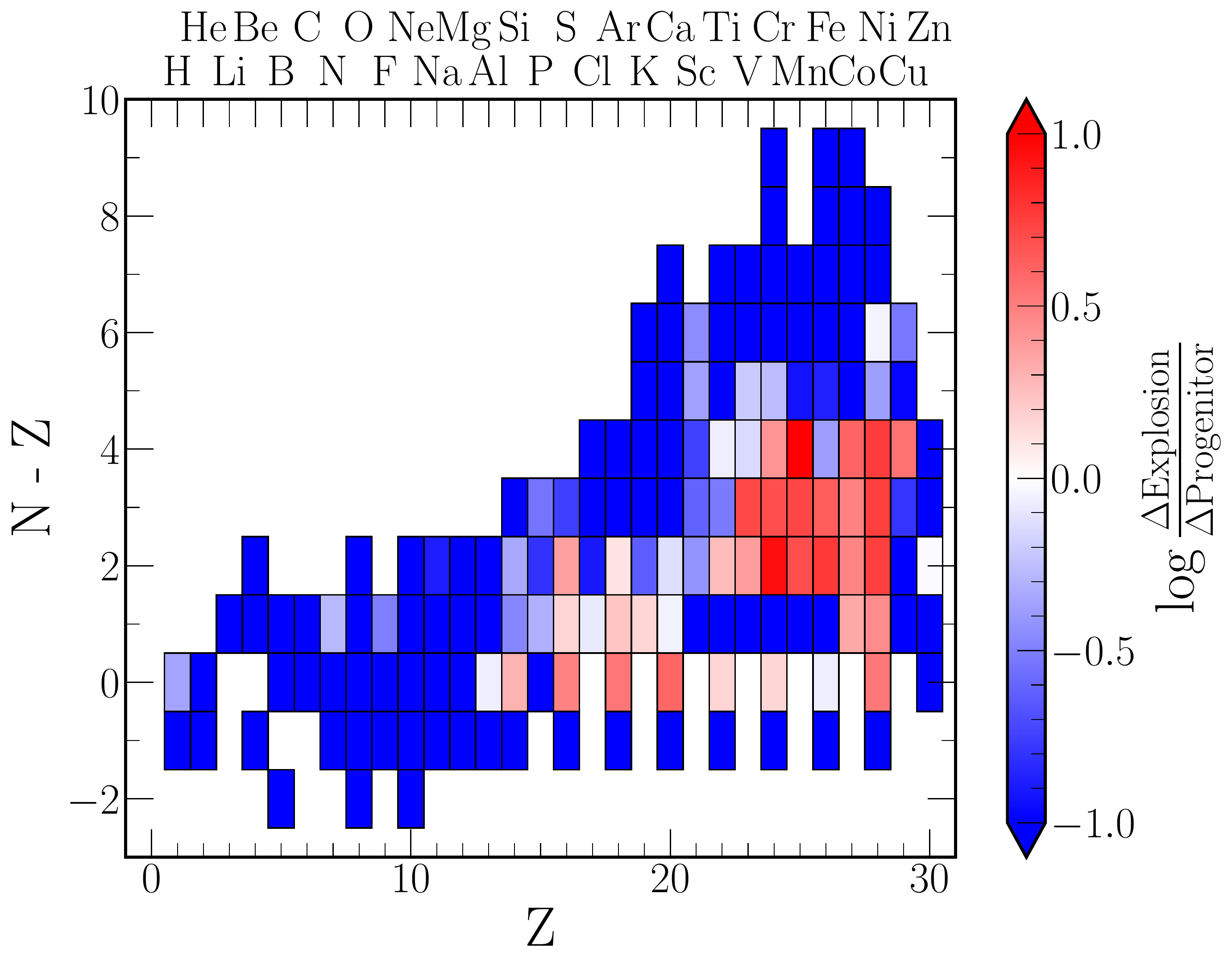}}
     \subfigure[Binary-stripped stars]{\label{fig:exp_binary}\includegraphics[width=0.49\linewidth]{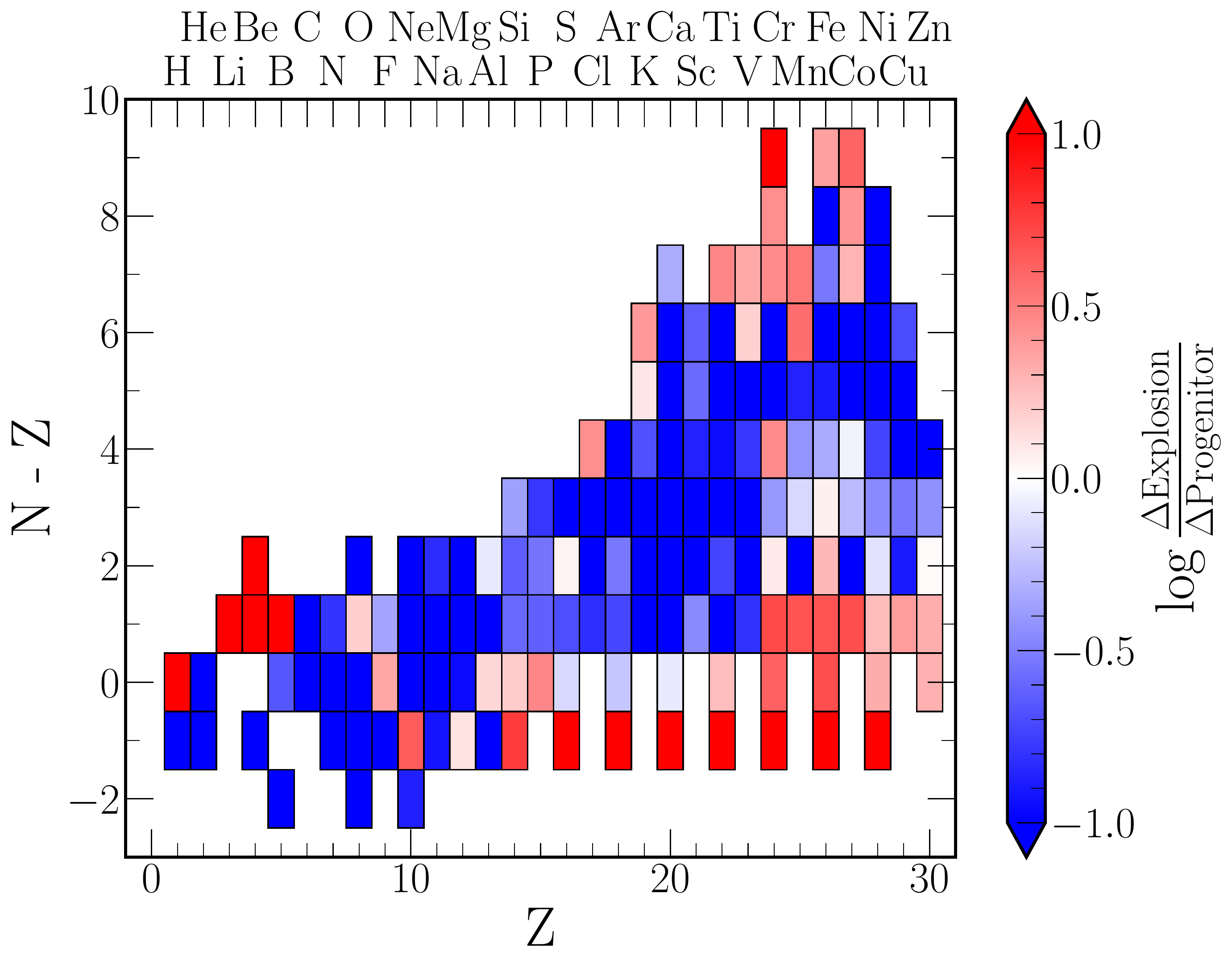}}
        \caption{ The logarithmic ratio of the variation in the isotopic ejecta when considering the uncertainties in explosion physics ($\Delta \rm{Explosion}$) and the uncertainties in the progenitors ($\Delta \rm {Progenitor}$).  $\Delta \rm{Explosion}$ is defined as the spread in the final ejecta composition for a 17 \msun{} model when the explosion physics is varied. $\Delta \rm {Progenitor}$ is the spread in the final ejecta composition for all models that explode with fixed explosion physics, varying the initial mass of the progenitor. The colors are truncated to be between -1 and 1.}
        \label{fig:explode_uncertainity}
\end{figure*}

Figure \ref{fig:explode_uncertainity} shows for which isotopes it is more important to understand the progenitor ($\log \left(\Delta \rm{Explosion} / \Delta \rm{Progenitor}\right) = -1.0$) and for which isotopes it is more important to understand the explosion mechanism ($\log \left(\Delta \rm{Explosion} / \Delta \rm{Progenitor}\right) > -1.0$ ).  Where $\Delta \rm{Explosion}$ is the spread in the final ejecta composition for a 17\msun{} model where we vary the explosion physics, and $\Delta \rm{Progenitor}$ is the spread in the final ejecta composition for fixed explosion physics but considering our entire range of initial masses. Both 
$\Delta \rm{Explosion}$ and $\Delta \rm{Progenitor}$ are defined separately for our single and binary-stripped stars.

Figure \ref{fig:explode_uncertainity} also shows that there is a difference between single stars and binary-stripped stars in which isotopes are uncertain, and thus which types of supernovae could be used to constrain different isotopes if detectable either through spectroscopy or radioactive decay signatures. These differences occur due to the differences in the time needed for the shock to break out of the photosphere, in the less massive single stars (which still have a RSG envelope) shock breakout is approximately a day after the core collapses, while in our binary-stripped stars (and high-mass single stars that lose their H-envelopes) shock breakout occurs much less than a day after the core collapses.
Thus stripped-envelope supernovae explode faster and have less time for short lived radioactive nuclei to decay before the shock breakout phase. This hints that stripped-envelope supernovae from binaries could be interesting locations to study radioactive decay products from explosive nucleosynthesis, with sufficiently early time observations and sufficient chemical mixing to bring them to the photosphere.

\subsection{Gamma ray emitters}

In \citet{andrews20} a number of isotopes (\potassium[43], \calcium[47], \scandium[44], \scandium[47], \vanadium[48], \chromium[48], \chromium[51], \manganese[52], \iron[59], \cobalt[56], \cobalt[57], and \nickel[57]) were determined to be interesting candidates for next-generation gamma-ray detectors. This was due to their 
accessible gamma-ray energies as well as their sensitivity to the structure of the
progenitor  (\vanadium[48] and \cobalt[57]) or the explosion energy (\potassium[43], \calcium[47], \scandium[44], \scandium[47], \iron[59]).

If we assume the (arbitrary) point of $\log \left(\Delta \rm{Explosion} / \Delta \rm{Progenitor}\right) > 0.1$ as the point where isotopes are probing for the explosion mechanism (instead of them
profiling the stellar structure) we can then use the isotope list of \citet{andrews20} to determine whether these isotopes are interesting for our binary-stripped stars.
We find that the following are interesting isotopes, by that definition, for probing the explosion properties in single stars: \vanadium[48], \chromium[48], \chromium[51], \manganese[52], \cobalt[56], \cobalt[57], and \nickel[57], while for binaries it is: \scandium[44], \chromium[48], \iron[59], \cobalt[57], and \nickel[57]. Note that the two lists do not overlap, nor does each isotope have the same level of sensitivity to the explosion between single and binary-stripped stars. We find \vanadium[48] and \manganese[52] are much stronger probes of the explosion physics in single stars than in binaries, while 
\chromium[48] is more sensitive to the explosion mechanism in binaries than in singles. 

\section{Caveats}\label{sec:caveats}

\subsection{Al26}\label{sec:al26}

\aluminum[26] is a short-lived isotope that decays to an excited state of \magnesium[26]. This 
excited \magnesium[26] then rapidly emits a 1.81 MeV $\gamma$-ray. Given the the short half life of \aluminum[26], its detection in the Galaxy requires a continuous production source \citep{diehl10}. \aluminum[26] has multiple formation sites inside stars, mainly via 
proton captures onto \magnesium[25] \citep{limongi06}. However, \aluminum[26] has several isomers which complicate the decay chain.
The ground state of \aluminum[26] decays to \magnesium[26] with a half life of $\sim700$ kyr, however
the excited state of \aluminum[26] decays to \magnesium[26] in $\sim6$ seconds \citep{misch21}.
At the temperatures found in stellar envelopes, excited \aluminum[26] decays to \magnesium[26] before it can de-excite back to its ground state, this rapidly converts most of the \aluminum[26] into \magnesium[26].
Our nuclear network does not 
contain the \aluminum[26] isomers, with the decay rate of \aluminum[26] only considering the ground state transitions. Thus we significantly over produce the amount of
\aluminum[26] in our wind yields and RLOF ejecta (see for instance the models of \citet{brinkman19,brinkman21} for \aluminum[26] yields from binary stars).
This over production ranges from a factor 10 to 10,000 depending on the initial mass, with low mass stars having larger overproduction factors.
Thus we can make no statement on the production of \aluminum[26] in stellar winds, without further post-processing of the chemical yields. Tests with a nuclear network that contains the \aluminum[26] isomers give wind and RLOF results that are consistent with those of \citet{brinkman21}. In our models we have \magnesium[26] wind yields of $M\approx10^{-3}\msun$, while \aluminum[26] wind yields of $M\approx10^{-4}\msun$ (at $\mint\approx20\msun$). Thus the lack of \aluminum[26] excited-state decays to \magnesium[26] leads to a systematic change of order 10\% (depending on the initial stellar mass) in the \magnesium[26] wind yields.

\subsection{Neutral ion electron captures}\label{sec:ignore_neut}

\MESA's default reaction rate for $\beryllium[7](e^{-},\nu_e)\lithium[7]$ (for version r12115)
came from \REACLIB{} \citep{cyburt10}. However, \REACLIB{} is only defined for temperatures above $T>10^{7}$K, thus it assumes that all atoms are ionised. While this is a reasonable assumption for most rates, which occur deep in the stellar interior, significant lithium
production occurs in stellar envelopes with temperatures below $T<10^{7}$K where the reaction rate can then depend on the level of ionisation \citep{schwab2020}. This has been corrected in \MESA{} version \texttt{2c2ad6c} by switching to the rate of \citet{simonucci13} which includes this effect, but this was not
available in \MESA{} for this work. We caution against making any inferences for either
\beryllium[7] or \lithium[7] from this work. Other isotopes that are sensitive to the level of ionisation may also be affected, if their rates are provided by \REACLIB{}.

\section{Discussion}\label{sec:discus}

We have used in this work the default \MESA{} set of reaction rates, which is a mix of rates from NACRE and JINA's REACLIB \citep{angulo99,cyburt10}. These rates in turn are based on experimental data and thus have experimental errors associated with them \citep{sallaska13}.  Changes in the nuclear reaction rates can have two different effects on the resulting nucleosynthesis, directly changing the amount of a species produced \citep{bliss20,subedi20,sieverding22}, and secondly indirectly by changing stellar properties like the core mass \citep{brown99,brown01,west13,fields18,sukhbold20,farmer20}. Of critical importance is the \crate{} reaction \citep{deboer17,farmer20} during core helium burning. Variations within its known uncertainties can lead to core masses changing by $\pm 2\msun$ \citep{sukhbold20} and central carbon fractions by $\pm80\%$ \citep{fields16,fields18}. This can alter models such that they change from exploding to non-exploding, significantly changing the final yields. Observations of WO stars suggest
that the \crate{} reaction needs a reduction by up to 50\% to match the observed C/O ratio \citep{aadland2022}. Further work is need to integrate reaction rate uncertainties, for all reaction rates, into stellar models.

Significant uncertainties remain in the treatment of mass transfer between binaries \citep[e.g.][]{klencki20,klencki21,sen22}, whether the stars lose their entire hydrogen envelopes \citep{gotberg17,yoon:17,laplace20}, and how much of the material lost by the donor is accreted by the companion \citep{demink:09,janssens21}. In this work we have assumed fully conservative mass transfer (all material lost is accreted by the companion), if mass transfer was not fully conservative (some mass is lost during RLOF) then the Universe can be enriched with those products lost from the hydrogen envelope (see Section \ref{sec:rlof}).

Related to whether a star fully strips during RLOF is the wind mass loss, and its scaling with the metallicity. Lower wind mass loss rates, either intrinsically or due to lower metallicities \citep{sander20a,sander20b}, may allow stars to keep part of their hydrogen envelopes \citep{gotberg17,yoon:17,laplace20}. If a star keeps its hydrogen envelope then the yields for the higher-mass elements seen in Figure \ref{fig:imf_all} will decrease as the star will not expose its helium core. At low metallicities, binary stars may not become fully stripped \citep{gotberg17,yoon:17,laplace20,klencki21}, this would lead to the binary stars having yields closer to those of single stars.

We modelled our binaries as Case B systems, where stable RLOF occurs during the Hertzsprung gap. Depending on the initial period, binaries can also interact earlier during the main sequence (Case A) or later during core helium burning (Case C). Interaction during Case A RLOF is likely to lead to a smaller core mass \citep{schneider21}, which would then be more likely to explode successfully in a SN. While interaction during Case C may occur too late in the star`s life to significantly affect the core evolution. Though, if the star was likely to form a BH, then Case C mass loss will liberate some of the envelope that would have otherwised collapsed into the BH. 

The consequences of stellar mergers for the final structure also depend on the timing of the merger.  Mergers which occur when both stars are on the main sequence (MS) are expected to increase the core mass of the resulting star, as compared to that of the primary star in the pre-merger binary, and the later evolution would likely be approximately described by that of an initially more massive single star (although see, e.g., \citealt{Renzo+2022accretors} for details of the structure which may also differ for post-merger stars).
For post-MS mergers, any erosion of the hydrogen-depleted core --- either during the merger \citep[e.g.][]{Ivanova+Podsiadlowski2003} or by dredge-out during post-merger relaxation \citep{Justham+14} --- would not be reversed by growth of a convective MS core. The consequences of this decrease in core mass may make many such merger products more explodable than the primary star would have become if single, despite the increase in stellar mass \citep{Justham+14}.

We have ignored the evolution of the accretor in this work. Their most significant contribution would likely be due to the mass gain, moving stars that are initially too light to explode ($\mint\approx6-8$\msun) to gain enough mass that they now explode \citep{podsiadlowski04,zapartas:17,zapartas19}. Secondly, their contribution will depend on how much of the mass that they accreted, remains bound to the accretor and is further processed inside the accretor. For instance, SN 2016bkv shows nucleosynthetic signals that it may have accreted material, but with no significant reprocessing of the accreted material \citep{deckers21}. Detailed modelling of the nucleosynthetic signature of accretors is deferred to later work.

We modelled our stars as non-rotating objects, however rotation can play a key role in the evolution of massive stars \citep{meynet00}. This is due to both changes in the core structure \citep{ekstrom12,murphy21}, chemical mixing of elements to the surface \citep{meynet06,groh19}, and changes to the observed chemical composition \citep{hunter07,hunter08,hunter09}. Indeed, rotation may be needed to explain s-process enrichment in the Solar system \citep{rizzuti19,rizzuti21,prantzos20}. Rotation will also play a role during the explosion, and whether the compact object forms a jet which may alter the final nucleosyntheic signal \citep{winteler12,reichert22}.

\section{Conclusion}\label{sec:conc}

Motivated by the differences we found in the \carbon{} yields in \citet{farmer21} we have extended our study of the nucleosynthesis of massive binary-stripped and single stars with a larger nuclear network. With the high fraction of massive stars being inferred to be in binaries, accurate modelling of binary evolution is needed to be able to predict the nucleosynthetic yields of massive stars.

To achieve this we have modelled the evolution of binary-stripped and single stars from the onset of hydrogen burning to core collapse, at solar metallicity. We have then modelled the nucleosynthetic yields due to the supernovae shock. This was done using a large 162-isotope nuclear network for all phases of the star's evolution.

Our results can be summarised as follows:

\begin{itemize}

\item We find that, in general, stars stripped in binaries synthesize and then eject more mass of all elements than single stars. The increase in the yields depends strongly on the isotope. This is due to the increased mass loss in binary-stripped stars ejecting more material during the star's evolution and binary-stripped stars being more explodable and thus ejecting more material during the supernovae. 

\item \carbon[13]/\carbon[12] from massive stars may provide information on the efficiency of mass transfer. This is because \carbon[13]{} is primarily only ejected during RLOF, while \carbon[12] is ejected primarily during WR wind mass loss.

\item In qualitative agreement with previous work, we find that stars stripped in binaries are somewhat more likely to successful explode in a core-collapse supernova, especially at higher masses. This is due the changes in the core mass due to the mass loss in RLOF.

\item Depending on the element, the difference between single and binary-stripped stars can be as large as the difference between different software instruments. However, we stress we have not attempted to calibrate our results to observations.

\item We have identified a number of radioactive isotopes that may prove useful in constraining the explosion mechanism or progenitor structure with next generation gamma-ray detectors.

\item In the online Zenodo material we provide tables of chemical yields for all isotopes for all types of mass loss.

\end{itemize}

We conclude that the yields from massive binary-stripped stars are different from those of massive single stars, and that this difference depends on the isotope. Thus there is no single scale factor for taking into the account of binary interactions on the chemical yields. Therefore the effects of binarity require explicit modelling of all isotopes that are of interest.

\appendix

\section{Other physics choices}\label{sec:other_phys}

\MESA{} depends on a number of sources for its micro-physics; with its EOS being a blend of several sources; OPAL \citep{Rogers2002}, SCVH
\citep{Saumon1995}, PTEH \citep{Pols1995}, HELM
\citep{Timmes2000}, and PC \citep{Potekhin2010}.
While opacities are primarily from OPAL \citep{Iglesias1993,
Iglesias1996} with additional data from \citet{Buchler1976,Ferguson2005,Cassisi2007}

The nuclear reaction rates are a combination of
NACRE and JINA's REACLIB \citep{angulo99,cyburt10} compilations. Additional weak reaction rates are from \citet{Fuller1985, Oda1994,
Langanke2000}. Screening of the nuclear reaction rates
is computed with the prescription of \citet{Chugunov2007}.  Thermal
neutrino loss rates are provided from the fits of \citet{Itoh1996}. The Roche lobe radii in binary systems is computed from the fit of \citet{Eggleton1983}.
In our Roche lobe overflowing binary systems we compute the mass loss rate using the 
the prescription of \citet{kolb1990}.

\section{Data tables}\label{sec:data_table}

Table \ref{tab:yields} shows the \fluorine[19] yields from our models. This includes both the wind mass loss, RLOF, and supernovae yields. We provide supernova yields for all models, as there are multiple different criteria that can be used to determine whether a star successfully ejects its envelope in a supernovae. The online Zenodo material contains yield tables for all 162-isotopes from our nuclear network.

\begin{deluxetable*}{c|cccc|cccccc}
\tablewidth{\linewidth}
\tablecaption{A representative sample of the data available. Data is available online for all 162 isotopes. The amount of \fluorine[19] ejected in solar masses broken down by mass-loss type for both single stars and the primary star in the binary. ``Init'' marks the amount that is from the initial composition of the star. All ejecta masses are in units of $10^{-5}$ solar masses. Blank values indicate models that did not evolve successfully to core collapse.  We have included the supernova yields from all models, assuming all models successfully explode.\label{tab:yields}  }

 \tablehead{
\colhead{Initial mass [\msun]} & \multicolumn{4}{c}{Single [$10^{-5}$\msun]} & \multicolumn{5}{c}{Binary [$10^{-5}$\msun]} \\
& $\rm{M_{Winds}} $ & $\rm{M_{Winds,init}}$ & $\rm{M_{CC}}$ & $\rm{M_{CC,init}}$ & $\rm{M_{Winds}} $ & $\rm{M_{Winds,init}} $ & $\rm{M_{RLOF}} $ & $\rm{M_{RLOF,init}} $ & $\rm{M_{CC}} $ & $\rm{M_{CC,init}} $ }
\startdata
11 & 0.051 & 0.058 & 1.262 & 0.276 & 0.009 & 0.025 & 0.209 & 0.251 & 0.265 & 0.064 \\
12 & 0.070 & 0.079 & 0.206 & 0.287 & 0.011 & 0.031 & 0.222 & 0.266 & 1.459 & 0.076 \\
13 & 0.092 & 0.104 & 1.045 & 0.299 & 0.013 & 0.037 & 0.234 & 0.280 & 1.248 & 0.089 \\
14 & 0.118 & 0.135 & 0.888 & 0.298 & 0.015 & 0.045 & 0.245 & 0.293 & 0.120 & 0.097 \\
15 & 0.149 & 0.170 & 0.256 & 0.293 & 0.017 & 0.054 & 0.255 & 0.306 & 0.622 & 0.111 \\
16 & 0.183 & 0.211 & 0.232 & 0.291 & 0.021 & 0.063 & 0.263 & 0.320 & 0.849 & 0.122 \\
17 & 0.210 & 0.245 & 0.266 & 0.300 & 0.024 & 0.075 & 0.271 & 0.330 & 0.238 & 0.135 \\
18 & 0.235 & 0.279 & 0.345 & 0.298 & 0.028 & 0.086 & 0.277 & 0.345 & 0.384 & 0.143 \\
19 & 0.244 & 0.298 & 0.475 & 0.318 & 0.033 & 0.096 & 0.282 & 0.363 & 0.794 & 0.152 \\
20 & 0.235 & 0.297 & 0.621 & 0.347 & 0.040 & 0.074 & 0.286 & 0.382 & 0.375 & 0.192 \\
21 & 0.243 & 0.315 & 0.642 & 0.363 & 0.047 & 0.090 & 0.284 & 0.391 & 0.432 & 0.199 \\
22 & 0.273 & 0.356 & 0.585 & 0.360 & 0.061 & 0.086 & 0.267 & 0.363 & 0.271 & 0.262 \\
23 & 0.323 & 0.421 & 1.268 & 0.310 & 0.068 & 0.095 & 0.275 & 0.381 & 0.240 & 0.264 \\
24 & 0.337 & 0.447 & 10.390 & 0.328 &  &  &  &  &  &  \\
25 & 0.331 & 0.434 & 1.472 & 0.378 &  &  &  &  &  &  \\
26 & 0.367 & 0.496 & 1.539 & 0.348 & 0.086 & 0.189 & 0.283 & 0.399 & 2.250 & 0.259 \\
27 & 0.374 & 0.514 & 1.001 & 0.365 & 0.110 & 0.170 & 0.248 & 0.355 & 1.323 & 0.357 \\
28 & 0.371 & 0.532 & 0.706 & 0.382 &  &  &  &  &  &  \\
29 & 0.382 & 0.550 & 0.732 & 0.402 &  &  &  &  &  &  \\
30 & 0.379 & 0.555 & 0.780 & 0.433 & 0.134 & 0.264 & 0.240 & 0.366 & 6.873 & 0.367 \\
31 & 0.395 & 0.575 & 0.853 & 0.445 & 0.147 & 0.238 & 0.248 & 0.373 & 0.972 & 0.413 \\
32 & 0.397 & 0.593 & 1.352 & 0.460 & 0.149 & 0.329 & 0.252 & 0.385 & 28.069 & 0.339 \\
33 & 0.408 & 0.650 & 1.943 & 0.435 & 0.264 & 0.355 & 0.257 & 0.386 & 1.573 & 0.346 \\
34 & 0.412 & 0.661 & 2.232 & 0.464 & 0.207 & 0.380 & 0.245 & 0.370 & 1.518 & 0.375 \\
35 & 0.419 & 0.695 & 32.298 & 0.468 & 0.678 & 0.409 & 0.240 & 0.385 & 1.527 & 0.366 \\
36 & 0.422 & 0.737 & 0.913 & 0.453 & 0.635 & 0.431 & 0.224 & 0.389 & 1.293 & 0.376 \\
37 & 0.461 & 0.749 & 1.396 & 0.476 & 1.343 & 0.477 & 0.228 & 0.366 & 1.398 & 0.391 \\
38 & 0.529 & 0.770 & 1.416 & 0.489 & 1.420 & 0.501 & 0.219 & 0.362 & 1.516 & 0.409 \\
39 & 0.624 & 0.793 & 1.415 & 0.500 & 1.519 & 0.529 & 0.208 & 0.357 & 1.499 & 0.423 \\
40 & 0.696 & 0.815 & 1.319 & 0.516 & 1.566 & 0.554 & 0.198 & 0.352 & 1.541 & 0.435 \\
41 & 0.792 & 0.838 & 1.250 & 0.544 & 1.628 & 0.581 & 0.188 & 0.347 & 1.432 & 0.446 \\
42 & 0.905 & 0.863 & 1.328 & 0.549 & 1.703 & 0.610 & 0.176 & 0.340 & 1.496 & 0.471 \\
43 & 0.994 & 0.886 & 1.243 & 0.561 & 1.743 & 0.638 & 0.163 & 0.333 & 1.217 & 0.467 \\
44 & 1.096 & 0.911 & 2.062 & 0.582 & 1.967 & 0.687 & 0.150 & 0.318 & 1.375 & 0.485 \\
45 & 1.175 & 0.934 & 1.293 & 0.591 & 1.998 & 0.715 & 0.137 & 0.310 & 1.211 & 0.504 
\enddata

 \end{deluxetable*}

\section{Wind sensitivity}\label{sec:winds_eta}

\begin{figure}[ht]
  \centering
  \includegraphics[width=1.0\linewidth]{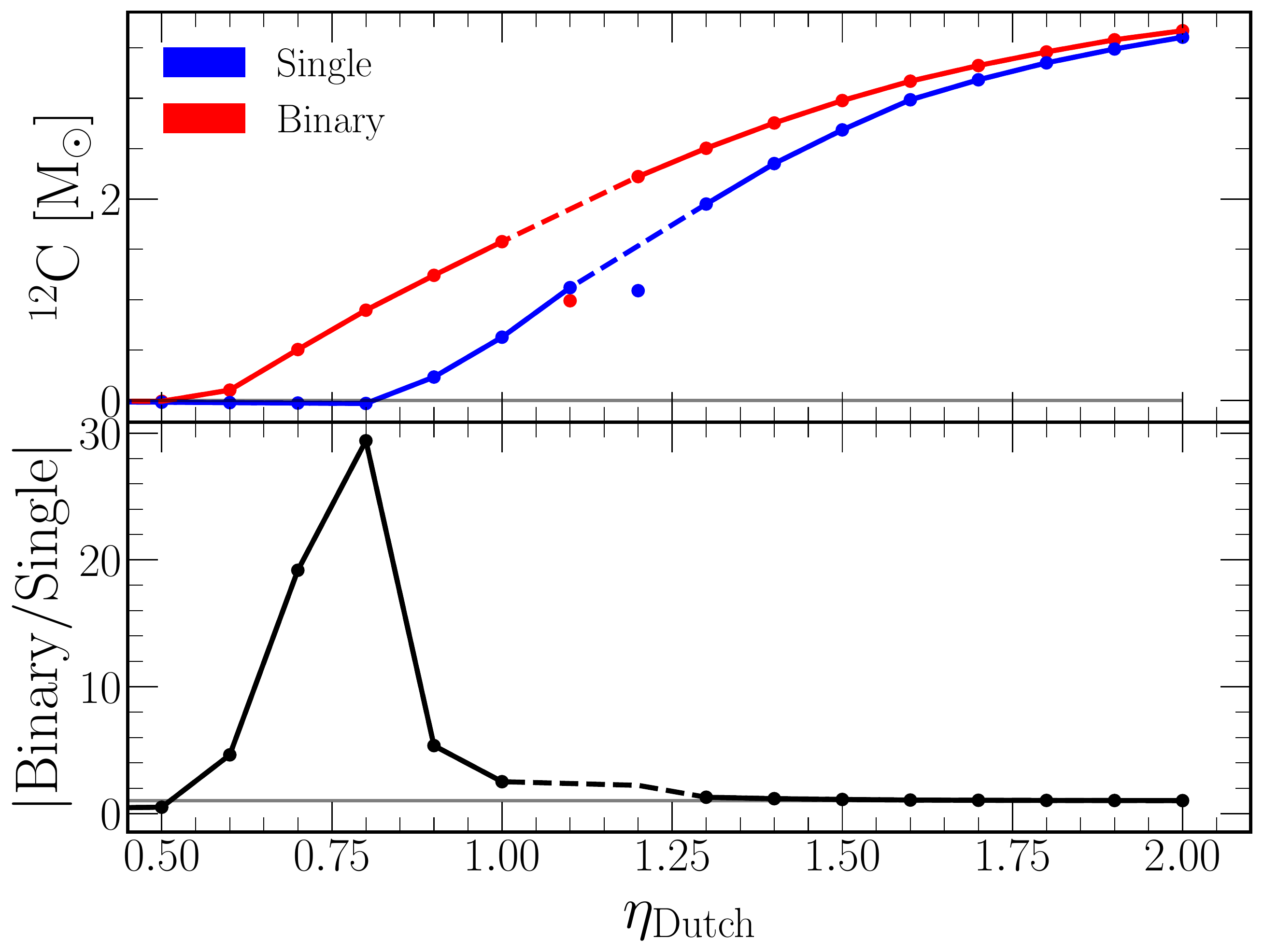}
  \caption{The top panel shows the \carbon[12] yield in solar masses for \mint=45\msun, single (blue) and binary (red), stars as a function of the 
  wind scaling factor $\rm{\eta_{Dutch}}$. The grey line in this panel shows a value of zero.
  The bottom panels shows the absolute values of the ratio of the two yields. Two models are shown that have anomalous behaviour and are excluded from the curves. The grey line in this panel shows where the ratio equals one.}
  \label{fig:winds}
\end{figure}

Figure \ref{fig:winds} shows the ratio of the \carbon[12] yields between a 45\msun{} binary-stripped and single star as a function of the wind scale factor $\rm{\eta_{Dutch}}$.
As $\eta_{Dutch}$ increases the total wind mass loss increases. This increases the amount of the C/O rich helium core removed once the star is stripped of its hydrogen and helium layers. For all $\rm{\eta_{Dutch}}> 0.5$ binary-stripped stars produce more \carbon[12] than single stars, with a peak in the fraction coming from binaries at $\rm{\eta_{Dutch}}\approx0.8$. This is due to the single star not removing its helium layers and thus not exposing its C/O rich helium core, while the binary star does \citep{farmer21}. The online Zenodo material contains the wind yields for all isotopes, for a $\mint=45\msun$ single and binary-stripped stars.

\section{Evolution properties}\label{sec:appen_evol}

Tables \ref{tab:evol_single} and \ref{tab:evol_bin} detail the main properties of our stellar models during their evolution.

\begin{deluxetable*}{c|cccccccccc}
\tablewidth{\linewidth}
\tablecaption{Properties of our single star models during their evolution.
$\rm{M_{init}}$ is the initial mass, $\rm{M_{Final}} $ is the final mass of the star, Age is the total lifetime of the star, $\rm{He_{core}}$ is the helium core mass at collapse, $\rm{C_{core}}$ is the carbon core mass at collapse, $\rm{\Delta M_{wind}}$ is the mass list due to winds, $\rm{\Delta M_{RLOF}}$ is the mass loss due to RLOF, $\rm{\carbon[12]_{cntr}}$ is the central \carbon[12] fraction at core helium depletion, $\rm{H_{fin}}$, $\rm{He_{fin}}$, $\rm{C_{fin}}$ are the final mass of \hydrogen, \helium, and \carbon{} at collapse.
\label{tab:evol_single}
}
 \tablehead{$\rm{M_{init}}$ & $\rm{M_{Final}} $ & Age & $\rm{He_{core}}$ & $\rm{C_{core}}$ & $\rm{\Delta M_{wind}}$ & $\rm{\Delta M_{RLOF}}$ & $\rm{\carbon[12]_{cntr}}$ & $\rm{H_{fin}}$ & $\rm{He_{fin}}$ & $\rm{C_{fin}}$ \\
\msun & \msun & Myr & \msun & \msun & \msun & \msun &  & \msun & \msun & \msun  }
\startdata
11 & 9.36 & 23.59 & 3.78 & 2.34 & 1.64 & 0.00 & 0.33 & 3.82 & 2.98 & 0.13 \\
12 & 9.77 & 20.38 & 4.27 & 2.75 & 2.23 & 0.00 & 0.32 & 3.75 & 3.04 & 0.12 \\
13 & 10.04 & 17.95 & 4.78 & 3.16 & 2.96 & 0.00 & 0.31 & 3.58 & 3.07 & 0.18 \\
14 & 10.18 & 16.06 & 5.29 & 3.61 & 3.82 & 0.00 & 0.30 & 3.30 & 3.05 & 0.18 \\
15 & 10.18 & 14.55 & 5.82 & 4.07 & 4.82 & 0.00 & 0.30 & 2.91 & 2.98 & 0.17 \\
16 & 10.03 & 13.32 & 6.36 & 4.54 & 5.97 & 0.00 & 0.29 & 2.41 & 2.86 & 0.18 \\
17 & 10.05 & 12.30 & 6.85 & 4.99 & 6.95 & 0.00 & 0.28 & 2.07 & 2.78 & 0.19 \\
18 & 10.08 & 11.45 & 7.34 & 5.44 & 7.92 & 0.00 & 0.28 & 1.74 & 2.70 & 0.31 \\
19 & 10.55 & 10.73 & 7.76 & 5.82 & 8.45 & 0.00 & 0.27 & 1.75 & 2.76 & 0.27 \\
20 & 11.57 & 10.11 & 8.04 & 6.08 & 8.43 & 0.00 & 0.27 & 2.22 & 3.05 & 0.32 \\
21 & 12.05 & 9.58 & 8.42 & 6.39 & 8.95 & 0.00 & 0.27 & 2.26 & 3.17 & 0.32 \\
22 & 11.90 & 9.12 & 8.78 & 6.74 & 10.10 & 0.00 & 0.26 & 1.81 & 3.13 & 0.34 \\
23 & 11.05 & 8.72 & 9.41 & 7.32 & 11.95 & 0.00 & 0.24 & 0.73 & 2.80 & 0.20 \\
24 & 11.31 & 8.35 & 9.15 & 7.81 & 12.69 & 0.00 & 0.23 & 0.57 & 2.75 & 0.30 \\
25 & 12.68 & 8.02 & 10.35 & 8.20 & 12.32 & 0.00 & 0.23 & 1.05 & 3.23 & 0.30 \\
26 & 11.94 & 7.72 & 11.70 & 9.47 & 14.06 & 0.00 & 0.21 & 0.05 & 2.20 & 0.37 \\
27 & 12.42 & 7.46 & 11.98 & 9.73 & 14.58 & 0.00 & 0.21 & 0.11 & 2.39 & 0.41 \\
28 & 12.89 & 7.23 & 12.15 & 9.87 & 15.11 & 0.00 & 0.21 & 0.23 & 2.60 & 0.39 \\
29 & 13.38 & 7.01 & 12.62 & 10.31 & 15.62 & 0.00 & 0.20 & 0.24 & 2.64 & 0.43 \\
30 & 14.26 & 6.80 & 13.09 & 10.73 & 15.74 & 0.00 & 0.20 & 0.40 & 2.92 & 0.31 \\
31 & 14.68 & 6.61 & 12.80 & 11.70 & 16.32 & 0.00 & 0.20 & 0.16 & 2.60 & 0.30 \\
32 & 15.16 & 6.44 & 14.41 & 11.98 & 16.84 & 0.00 & 0.20 & 0.23 & 2.72 & 0.24 \\
33 & 14.57 & 6.29 & 14.57 & 12.66 & 18.43 & 0.00 & 0.20 & 0.00 & 1.56 & 0.57 \\
34 & 15.24 & 6.14 & 15.24 & 13.60 & 18.76 & 0.00 & 0.20 & 0.00 & 1.59 & 0.64 \\
35 & 15.27 & 6.01 & 15.27 & \tablenotemark{a} & 19.73 & 0.00 & 0.20 & 0.00 & 1.18 & 0.91 \\
36 & 15.11 & 5.88 & 15.11 & \tablenotemark{a} & 20.89 & 0.00 & 0.20 & 0.00 & 0.82 & 0.66 \\
37 & 15.75 & 5.74 & 15.75 & 13.74 & 21.25 & 0.00 & 0.20 & 0.00 & 0.14 & 1.11 \\
38 & 16.16 & 5.63 & 16.16 & 12.44 & 21.84 & 0.00 & 0.20 & 0.00 & 0.13 & 1.20 \\
39 & 16.51 & 5.53 & 16.51 & 10.74 & 22.49 & 0.00 & 0.20 & 0.00 & 0.15 & 1.20 \\
40 & 16.89 & 5.44 & 16.89 & 14.23 & 23.11 & 0.00 & 0.20 & 0.00 & 0.28 & 1.19 \\
41 & 17.23 & 5.35 & 17.23 & 14.82 & 23.77 & 0.00 & 0.20 & 0.00 & 0.19 & 1.21 \\
42 & 17.53 & 5.26 & 17.53 & 14.31 & 24.47 & 0.00 & 0.19 & 0.00 & 0.21 & 1.28 \\
43 & 17.86 & 5.18 & 17.86 & 9.57 & 25.14 & 0.00 & 0.19 & 0.00 & 0.19 & 1.24 \\
44 & 18.16 & 5.10 & 18.16 & 7.30 & 25.84 & 0.00 & 0.19 & 0.00 & 0.22 & 2.00 \\
45 & 18.50 & 5.03 & 18.50 & 15.84 & 26.50 & 0.00 & 0.19 & 0.00 & 0.14 & 1.35 
\enddata
\tablenotetext{a}{The C core mass is ill-defined due to convective mixing bluring the chemical gradient at the edge of the C core in these models. }
\end{deluxetable*}

\begin{deluxetable*}{c|cccccccccc}
\tablewidth{\linewidth}
\tablecaption{Properties of our binary-stripped stars during their evolution.
Columns have the same meaning as in Table \ref{tab:evol_single}.
\label{tab:evol_bin}
}
 \tablehead{$\rm{M_{init}}$ & $\rm{M_{Final}} $ & Age & $\rm{He_{core}}$ & $\rm{C_{core}}$ & $\rm{\Delta M_{wind}}$ & $\rm{\Delta M_{RLOF}}$ & $\rm{\carbon[12]_{cntr}}$ & $\rm{H_{fin}}$ & $\rm{He_{fin}}$ & $\rm{C_{fin}}$ \\
\msun & \msun & Myr & \msun & \msun & \msun & \msun &  & \msun & \msun & \msun  }
\startdata
11 & 3.17 & 23.69 & 3.17 & 1.93 & 0.70 & 7.12 & 0.38 & 0.00 & 1.16 & 0.06 \\
12 & 3.58 & 20.44 & 3.58 & 2.22 & 0.87 & 7.55 & 0.37 & 0.00 & 1.24 & 0.08 \\
13 & 3.98 & 17.99 & 3.98 & 2.55 & 1.06 & 7.95 & 0.36 & 0.00 & 1.28 & 0.16 \\
14 & 4.39 & 16.08 & 4.39 & 2.88 & 1.29 & 8.32 & 0.35 & 0.00 & 1.35 & 0.11 \\
15 & 4.79 & 14.56 & 4.79 & 3.23 & 1.54 & 8.67 & 0.34 & 0.00 & 1.33 & 0.19 \\
16 & 5.15 & 13.33 & 5.15 & 3.56 & 1.78 & 9.07 & 0.33 & 0.00 & 1.34 & 0.22 \\
17 & 5.52 & 12.31 & 5.52 & 3.91 & 2.13 & 9.35 & 0.32 & 0.00 & 1.31 & 0.22 \\
18 & 5.77 & 11.46 & 5.77 & 4.14 & 2.43 & 9.80 & 0.32 & 0.00 & 1.31 & 0.25 \\
19 & 6.00 & 10.75 & 6.00 & 4.36 & 2.71 & 10.29 & 0.31 & 0.00 & 1.19 & 0.35 \\
20 & 7.07 & 10.12 & 7.07 & 5.27 & 2.09 & 10.83 & 0.29 & 0.00 & 1.39 & 0.36 \\
21 & 7.36 & 9.58 & 7.36 & 5.55 & 2.56 & 11.08 & 0.29 & 0.00 & 1.36 & 0.35 \\
22 & 9.27 & 9.15 & 8.48 & 6.49 & 2.43 & 10.30 & 0.25 & 0.32 & 2.31 & 0.35 \\
23 & 9.50 & 8.71 & 7.44 & 7.09 & 2.70 & 10.80 & 0.26 & 0.14 & 2.12 & 0.40 \\
24 &  &  &  &  &  &  &  &  &  &  \\
25 &  &  &  &  &  &  &  &  &  &  \\
26 & 9.31 & 7.70 & 9.31 & 7.46 & 5.37 & 11.32 & 0.27 & 0.00 & 0.86 & 0.84 \\
27 & 12.12 & 7.47 & 11.15 & 8.97 & 4.81 & 10.07 & 0.23 & 0.36 & 2.58 & 0.33 \\
28 &  &  &  &  &  &  &  &  &  &  \\
29 &  &  &  &  &  &  &  &  &  &  \\
30 & 12.12 & 6.79 & 12.12 & 0.00 & 7.50 & 10.38 & 0.23 & 0.00 & 1.33 & 0.72 \\
31 & 13.66 & 6.61 & 12.43 & 11.15 & 6.75 & 10.58 & 0.21 & 0.03 & 2.26 & 0.21 \\
32 & 11.75 & 6.43 & 11.75 & 9.72 & 9.32 & 10.93 & 0.24 & 0.00 & 0.38 & 1.04 \\
33 & 11.98 & 6.28 & 11.98 & 9.91 & 10.08 & 10.94 & 0.23 & 0.00 & 0.29 & 1.14 \\
34 & 12.71 & 6.14 & 12.71 & 10.53 & 10.78 & 10.51 & 0.23 & 0.00 & 0.27 & 1.22 \\
35 & 12.47 & 6.00 & 12.47 & 10.32 & 11.60 & 10.93 & 0.23 & 0.00 & 0.30 & 1.22 \\
36 & 12.73 & 5.88 & 12.73 & 10.55 & 12.24 & 11.04 & 0.23 & 0.00 & 0.29 & 1.25 \\
37 & 13.11 & 5.75 & 13.11 & 10.89 & 13.52 & 10.37 & 0.22 & 0.00 & 0.28 & 1.29 \\
38 & 13.50 & 5.64 & 13.50 & 6.22 & 14.22 & 10.28 & 0.22 & 0.00 & 0.17 & 1.12 \\
39 & 13.86 & 5.54 & 13.86 & 6.31 & 15.00 & 10.14 & 0.22 & 0.00 & 0.20 & 1.19 \\
40 & 14.29 & 5.44 & 14.29 & 6.55 & 15.72 & 9.99 & 0.21 & 0.00 & 0.20 & 1.23 \\
41 & 14.69 & 5.35 & 14.69 & 7.22 & 16.48 & 9.83 & 0.21 & 0.00 & 0.19 & 1.21 \\
42 & 15.06 & 5.27 & 15.06 & 6.49 & 17.29 & 9.65 & 0.21 & 0.00 & 0.19 & 1.28 \\
43 & 15.48 & 5.19 & 15.48 & 12.92 & 18.09 & 9.43 & 0.21 & 0.00 & 0.30 & 1.21 \\
44 & 15.49 & 5.11 & 15.49 & 7.01 & 19.49 & 9.01 & 0.21 & 0.00 & 0.18 & 1.22 \\
45 & 15.91 & 5.04 & 15.91 & 9.77 & 20.29 & 8.80 & 0.20 & 0.00 & 0.21 & 1.28 \\
\enddata
\end{deluxetable*}

\section{Explosion properties}

Tables \ref{tab:exp_single} and \ref{tab:exp_binary} detail the main properties of our stellar models during the core collapse phase of their evolution.

\begin{deluxetable*}{c|ccccccccc}
\tablewidth{\linewidth}
\tablecaption{Properties of our single star models during the supernovae explosion.
$\rm{M_{init}}$ is the initial mass, $\rm{M_{Final}} $ is the final mass of the star, $\rm{M_{Fe}}$ is the mass of the iron core, $\rm{M_{Ni56}} $ is the mass of \nickel[56] at the time of shock breakout, $\rm{E}$ is the total energy of the model at shock breakout, $\rm{M_{rem}}$ is the mass of the compact object at shock breakout (assuming the envelope did not fallback onto the compact object), $\rm{Fate}$
is the final fate as given by \citet{ertl:15}.
\label{tab:exp_single}
}

 \tablehead{$\rm{M_{init}}$ & $\rm{M_{Final}} $ & $\rm{M_{Fe}}$ & $\rm{\xi_{M=2.5}}$ & $\rm{M_{4}}$ & $\rm{\mu_{4}} $ & $\rm{M_{Ni56}} $ & $\rm{E} $ & $\rm{M_{rem}} $ & $\rm{Fate}$ \\
\msun & \msun & \msun & & \msun & $\msun / 1000\rm{km}$ & \msun & $10^{51} \rm{erg/s}$ & \msun & }
\startdata
11 &  9.36 & 1.36 & 0.05 & 1.52 & 0.06 & 0.04 & 1.00 & 1.53 & NS \\
12 &  9.77 & 1.35 & 0.09 & 1.62 & 0.05 & 0.02 & 1.00 & 1.62 & NS \\
13 &  10.04 & 1.40 & 0.18 & 1.57 & 0.10 & 0.08 & 1.00 & 1.57 & BH \\
14 &  10.18 & 1.43 & 0.22 & 1.73 & 0.10 & 0.04 & 1.00 & 1.73 & BH \\
15 &  10.18 & 1.47 & 0.19 & 1.88 & 0.07 & 0.04 & 1.00 & 1.88 & NS \\
16 &  10.03 & 1.47 & 0.22 & 1.77 & 0.10 & 0.06 & 1.00 & 1.78 & BH \\
17 &  10.05 & 1.41 & 0.17 & 1.56 & 0.12 & 0.14 & 1.00 & 1.56 & BH \\
18 &  10.08 & 1.39 & 0.19 & 1.62 & 0.08 & 0.04 & 1.00 & 1.62 & BH \\
19 &  10.55 & 1.40 & 0.18 & 1.53 & 0.10 & 0.11 & 1.00 & 1.53 & BH \\
20 &  11.57 & 1.44 & 0.21 & 1.71 & 0.09 & 0.06 & 1.00 & 1.71 & BH \\
21 &  12.05 & 1.41 & 0.16 & 1.75 & 0.06 & 0.02 & 1.00 & 1.75 & NS \\
22 &  11.90 & 1.43 & 0.15 & 1.68 & 0.08 & 0.08 & 1.00 & 1.68 & BH \\
23 &  11.05 & 1.64 & 0.41 & 2.27 & 0.12 & 0.07 & 1.00 & 2.27 & BH \\
24 &  11.31 & 1.70 & 0.51 & 2.02 & 0.27 & 0.31 & 1.00 & 2.02 & BH \\
25 &  12.68 & 1.71 & 0.53 & 1.96 & 0.30 & 0.39 & 1.00 & 1.96 & BH \\
26 &  11.94 & 1.61 & 0.39 & 2.08 & 0.14 & 0.14 & 1.00 & 2.08 & BH \\
27 &  12.42 & 1.62 & 0.38 & 2.06 & 0.14 & 0.13 & 1.00 & 2.06 & BH \\
28 &  12.89 & 1.62 & 0.38 & 2.06 & 0.14 & 0.13 & 1.00 & 2.06 & BH \\
29 &  13.38 & 1.57 & 0.34 & 1.98 & 0.12 & 0.13 & 1.00 & 1.98 & BH \\
30 &  14.26 & 1.49 & 0.26 & 1.99 & 0.09 & 0.06 & 1.00 & 1.99 & NS \\
31 &  14.68 & 1.51 & 0.25 & 2.07 & 0.07 & 0.06 & 1.00 & 2.07 & NS \\
32 &  15.16 & 1.53 & 0.26 & 2.12 & 0.07 & 0.05 & 1.00 & 2.12 & NS \\
33 &  14.57 & 1.65 & 0.44 & 2.22 & 0.15 & 0.17 & 1.00 & 2.22 & BH \\
34 &  15.24 & 1.67 & 0.49 & 2.09 & 0.25 & 0.30 & 1.00 & 2.09 & BH \\
35 &  15.27 & 1.51 & 0.29 & 2.00 & 0.13 & 0.12 & 1.00 & 2.00 & BH \\
36 &  15.11 & 1.60 & 0.37 & 2.25 & 0.10 & 0.08 & 1.00 & 2.25 & NS \\
37 &  15.75 & 1.57 & 0.33 & 2.24 & 0.08 & 0.07 & 1.00 & 2.24 & NS \\
38 &  16.16 & 1.58 & 0.34 & 2.28 & 0.08 & 0.05 & 1.00 & 2.28 & NS \\
39 &  16.51 & 1.58 & 0.37 & 2.32 & 0.08 & 0.06 & 1.00 & 2.32 & NS \\
40 &  16.89 & 1.61 & 0.44 & 2.26 & 0.15 & 0.14 & 1.01 & 2.26 & BH \\
41 &  17.23 & 1.57 & 0.40 & 1.79 & 0.21 & 0.33 & 1.00 & 1.79 & BH \\
42 &  17.53 & 1.59 & 0.42 & 1.96 & 0.20 & 0.26 & 1.00 & 1.96 & BH \\
43 &  17.86 & 1.58 & 0.43 & 1.94 & 0.22 & 0.28 & 1.00 & 1.94 & BH \\
44 &  18.16 & 1.49 & 0.33 & 1.66 & 0.19 & 0.30 & 1.00 & 1.66 & BH \\
45 &  18.50 & 1.55 & 0.38 & 1.74 & 0.19 & 0.33 & 1.00 & 1.74 & BH
\enddata
\end{deluxetable*}

\begin{deluxetable*}{c|ccccccccc}
\tablewidth{\linewidth}
\tablecaption{Properties of our binary-stripped models during the supernovae explosion.
Columns have the same meaning as in Table \ref{tab:exp_single}. \label{tab:exp_binary}
}
 \tablehead{$\rm{M_{init}}$ & $\rm{M_{Final}} $ & $\rm{M_{Fe}}$ & $\rm{\xi_{M=2.5}}$ & $\rm{M_{4}}$ & $\rm{\mu_{4}} $ & $\rm{M_{Ni56}} $ & $\rm{E} $ & $\rm{M_{rem}} $ & $\rm{Fate}$ \\
\msun & \msun & \msun & & \msun &  & \msun & $10^{51} erg/s$ & \msun &}
\startdata
11 &  3.17 & 1.26 & 0.01 & 1.36 & 0.03 & 0.03 & 1.00 & 1.36 & NS \\
12 &  3.58 & 1.32 & 0.03 & 1.43 & 0.06 & 0.08 & 1.00 & 1.43 & NS \\
13 &  3.98 & 1.32 & 0.10 & 1.46 & 0.06 & 0.07 & 1.00 & 1.46 & NS \\
14 &  4.39 & 1.35 & 0.10 & 1.64 & 0.04 & 0.02 & 1.00 & 1.64 & NS \\
15 &  4.79 & 1.40 & 0.18 & 1.64 & 0.08 & 0.08 & 1.00 & 1.64 & BH \\
16 &  5.15 & 1.46 & 0.21 & 1.69 & 0.10 & 0.12 & 1.00 & 1.69 & BH \\
17 &  5.52 & 1.44 & 0.17 & 1.69 & 0.08 & 0.11 & 1.00 & 1.69 & BH \\
18 &  5.77 & 1.44 & 0.19 & 1.71 & 0.09 & 0.11 & 1.00 & 1.71 & BH \\
19 &  6.00 & 1.43 & 0.19 & 1.69 & 0.09 & 0.10 & 1.00 & 1.69 & BH \\
20 &  7.07 & 1.38 & 0.21 & 1.62 & 0.10 & 0.11 & 1.00 & 1.62 & BH \\
21 &  7.36 & 1.44 & 0.20 & 1.70 & 0.09 & 0.10 & 1.00 & 1.70 & BH \\
22 &  9.27 & 1.46 & 0.17 & 1.85 & 0.06 & 0.06 & 1.00 & 1.85 & NS \\
23 &  9.50 & 1.51 & 0.23 & 2.03 & 0.08 & 0.06 & 1.00 & 2.03 & NS \\
24 &  &  &  &  &  &  &  &  & \\
25 &  &  &  &  &  &  &  &  & \\
26 &  9.31 & 1.58 & 0.34 & 1.96 & 0.15 & 0.16 & 1.00 & 1.96 & BH \\
27 &  12.12 & 1.69 & 0.50 & 1.99 & 0.26 & 0.34 & 1.00 & 1.99 & BH \\
28 &  &  &  &  &  &  &  &  & \\
29 &  &  &  &  &  &  &  &  & \\
30 &  12.12 & 1.38 & 0.13 & 1.71 & 0.05 & 0.03 & 1.00 & 1.71 & NS \\
31 &  13.66 & 1.43 & 0.18 & 1.95 & 0.06 & 0.02 & 1.00 & 1.95 & NS \\
32 &  11.75 & 1.69 & 0.49 & 2.13 & 0.22 & 0.26 & 1.01 & 2.13 & BH \\
33 &  11.98 & 1.65 & 0.46 & 2.15 & 0.19 & 0.21 & 1.00 & 2.15 & BH \\
34 &  12.71 & 1.62 & 0.39 & 2.08 & 0.14 & 0.15 & 1.00 & 2.08 & BH \\
35 &  12.47 & 1.62 & 0.39 & 2.09 & 0.14 & 0.15 & 1.00 & 2.09 & BH \\
36 &  12.73 & 1.61 & 0.38 & 2.06 & 0.14 & 0.15 & 1.00 & 2.06 & BH \\
37 &  13.11 & 1.60 & 0.36 & 2.03 & 0.13 & 0.14 & 1.00 & 2.03 & BH \\
38 &  13.50 & 1.45 & 0.18 & 1.90 & 0.06 & 0.06 & 1.00 & 1.90 & NS \\
39 &  13.86 & 1.46 & 0.19 & 1.87 & 0.08 & 0.10 & 1.00 & 1.87 & NS \\
40 &  14.29 & 1.48 & 0.20 & 1.94 & 0.07 & 0.08 & 1.00 & 1.94 & NS \\
41 &  14.69 & 1.49 & 0.22 & 2.03 & 0.07 & 0.07 & 1.00 & 2.03 & NS \\
42 &  15.06 & 1.48 & 0.21 & 1.69 & 0.14 & 0.20 & 1.00 & 1.69 & BH \\
43 &  15.48 & 1.58 & 0.32 & 2.24 & 0.08 & 0.05 & 1.00 & 2.24 & NS \\
44 &  15.49 & 1.51 & 0.23 & 1.73 & 0.14 & 0.20 & 1.00 & 1.73 & BH \\
45 &  15.91 & 1.47 & 0.28 & 1.60 & 0.19 & 0.32 & 1.00 & 1.60 & BH
\enddata
 \end{deluxetable*}

\clearpage

\acknowledgments

We acknowledge helpful discussions with M.\ Renzo.
SdM and SJ acknowledge funding by the Netherlands Organization for
Scientific Research (NWO) as part of the Vidi research program BinWaves
(project number 639.042.728). 
This work was also supported by the Cost Action Program ChETEC CA16117. 
EL acknowledges funding from the European Research Council (ERC) under the European Union’s Horizon 2020 research and innovation programme (Grant agreement No.\ 945806).
This research has made use of NASA's Astrophysics Data System.

\newpage
 \software{
 \texttt{mesaPlot} \citep{mesaplot},
 \texttt{mesaSDK} \citep{mesasdk},
 \texttt{ipython/jupyter} \citep{perez_2007_aa,kluyver_2016_aa},
 \texttt{matplotlib} \citep{hunter_2007_aa},
 \texttt{NumPy} \citep{der_walt_2011_aa},
 \texttt{Parallel} \citep{tange_ole_2021_5797028},
 \texttt{VICE} \citep{johnson20,johnson21,griffith21},
 \MESA \citep{paxton:11,paxton:13,paxton:15,paxton:18,paxton:19,jermyn2022},
 \texttt{pyMesa} \citep{pymesa}, and 
 \texttt{Webplotdigitizer} \citep{Rohatgi2022}.
          }
          
\newpage
\newpage

\bibliographystyle{aasjournal}
\bibliography{ccsn}

\end{document}